\documentclass[pre,superscriptaddress,twocolumn,showpacs]{revtex4-1}
\usepackage{graphicx,bbm}
\usepackage{hyperref}
\usepackage{amsmath}
\usepackage{float}
\def\etal#1{ {\em et al.}}
\def\tit#1{}
\def\ii{{\rm i}}

\makeatletter
\def\underbracket{\@ifnextchar [ {\@underbracket} {\@underbracket [\@bracketheight]}}
\def\@underbracket[#1]{\@ifnextchar [ {\@under@bracket[#1]} {\@under@bracket[#1][0.4em]}}
\def\@under@bracket[#1][#2]#3{
           \mathop {\vtop {\m@th \ialign {##\crcr $\hfil \displaystyle {#3}\hfil $%
                              \crcr \noalign {\kern 3\p@ \nointerlineskip }\upbracketfill {#1}{#2}
                              \crcr \noalign {\kern 3\p@ }}}}\limits}
\def\upbracketfill#1#2{$\m@th \setbox \z@ \hbox {$\braceld$}
                  \edef\@bracketheight{\the\ht\z@}\bracketend{#1}{#2}
                  \leaders \vrule \@height #1 \@depth \z@ \hfill
                  \leaders \vrule \@height #1 \@depth \z@ \hfill \bracketend{#1}{#2}$}
\def\bracketend#1#2{\vrule height #2 width #1\relax}
\def\downbracketfill#1#2{$\m@th \setbox \z@ \hbox {$\braceld$}
                  \edef\@bracketheight{\the\ht\z@}\downbracketend{#1}{#2}
                  \leaders \vrule \@height #1 \@depth \z@ \hfill
                  \leaders \vrule \@height #1 \@depth \z@ \hfill
\downbracketend{#1}{#2}$}
\def\downbracketend#1#2{\vrule depth #2 width #1\relax}
\makeatother

\begin{document}
\pagenumbering{arabic}

\title{Spectral Domain of Large Nonsymmetric correlated Wishart Matrices} 
\author{Vinayak}
\email{vinayaksps2003@gmail.com}
\affiliation{Instituto de Ciencias F\' isicas, Universidad Nacional Aut\' onoma de M\' exico, C.P. 62210 Cuernavaca, M\' exico}
\author{Luis Benet}
\email{benet@fis.unam.mx}
\affiliation  {Instituto de Ciencias F\' isicas, Universidad Nacional Aut\' onoma de M\' exico, C.P. 62210 Cuernavaca, M\' exico}
\affiliation {Centro Internacional de Ciencias, C.P. 62210 Cuernavaca, M\' exico}

\begin{abstract}
We study complex eigenvalues of the Wishart model for nonsymmetric correlation matrices. The model is defined for two statistically equivalent but different Gaussian real matrices, as $\mathsf{C}=\mathsf{AB}^{t}/T$, where $\mathsf{B}^{t}$ is the transpose of $\mathsf{B}$ and both matrices $\mathsf{A}$ and $\mathsf{B}$ are of dimensions $N\times T$. If $\mathsf{A}$ and $\mathsf{B}$ are uncorrelated, or equivalently if $\mathsf{C}$ vanishes on average, it is known that at large matrix dimension the domain of the eigenvalues of $\mathsf{C}$ is a circle centered-at-origin and the eigenvalue density depends only on the radial distances. We consider {\it actual} correlation in $\mathsf{A}$ and $\mathsf{B}$ and derive a result for the contour describing the domain of the bulk of the eigenvalues of $\mathsf{C}$ in the limit of large $N$ and $T$ where the ratio $N/T$ is finite. In particular, we show that the eigenvalue domain is sensitive to the correlations. For example, when $\mathsf{C}$ is diagonal on average with the same element $c\ne0$, the contour is no longer a circle centered at origin but a shifted ellipse. In this case, we explicitly derive a result for the spectral density which again depends only on the radial distances. For more general cases, we show that the contour depends on the symmetric and anti-symmetric parts of the correlation matrix resulting from the ensemble averaged $\mathsf{C}$. If the correlation matrix is normal then the contour depends only on its spectrum. We also provide numerics to justify our analytics.
\end{abstract}

\maketitle
\renewcommand*\thesection{\Roman{section}}
\renewcommand*\thesubsection{\thesection.\Roman{subsection}}

\section{Introduction}

Correlation matrices are fundamental in multivariate analysis \cite{bachelier,Wilks, Wishart, Muirhead}. Examples thereof are not only in econophyscis \cite{Finance1, Finance2, Finance3, Finance4, Thomas2012} but also in biological sciences \cite{gene,Seba:03} and atmospheric science \cite{diverseAT} {among others}. In such studies, random matrix theory (RMT) proved to be remarkably useful. The basic random matrix model for the symmetric correlation matrices is due to Wishart \cite{Wishart}, which {is} often being credited as the very origin of RMT \cite{Mehta,Muirhead}. Notably, beyond the conventional theme, potential of this model has been realized in physics \cite{Mehta,Brody81,GuhrGW98, BenRMP97, Verbaarschot} and in communication engineering \cite{Muller:Review} as well.

In multivariate analysis the Wishart model provides a framework against which the non-noisy correlations must be viewed \cite{marchenko,Finance1, Finance2, Finance3, Finance4, vrt2013,gene,Seba:03,diverseAT}. Nevertheless, a simple generalization by including the cross-correlations in this model improves the so-called {\it null hypothesis} \cite{vrt2013} and provides a better platform to understand the underlying correlations \cite{Silverstien,SenM,Burda:2005,Baik:2005,MousSimon,vp2010,guhr1}. These models are often referred to as the correlated Wishart orthogonal ensembles (CWOE). 

Recently, a Wishart model for nonsysmmetric correlation matrices \cite{John} has drawn considerable attention in quantitative finance \cite{Biely:08, Kwapien:2006, Bouchaud:2007, Bouchaud:2009, Stanley:2011, Livan:2012} and in visual and auditory cortex analysis \cite{Dorzdz,PRDorzdz,Kwapien:2000}. Motivations behind studying such nonsymmetric matrices are twofold: The nonsymmetry is natural for a correlation matrix describing statistics between two different statistical systems while the time-lagged correlation matrix, which is interesting from the viewpoint of forecasting models, is also nonsymmetric. {Time-lagged correlation matrices are important in applications and their eigenvalues statistics may lead to substantial clues about the nonsymmetry associated with the system. This motivated spectral analysis of nonsymmetric correlation matrices of several complex systems \cite{Biely:08, Kwapien:2006,PRDorzdz,Kwapien:2000} where the RMT results for the uncorrelated matrices \cite{Burda:2010,Kanzieper-Singh, Proof1,Proof2, Mario:2014} have been useful to mark the predominance of the symmetric or anti-symmetric part of the correlations matrix. Numerical analysis with this approach have been quite illustrative \cite{Biely:08,PRDorzdz} and therefore analytical results for the spectral statistics of the random matrix model, as used in Ref. \cite{vin2013}, become important.}

The Wishart model for nonsymmetric correlation matrices is defined via two statistically equivalent but different Gaussian real matrices $\mathsf{A}$ and $\mathsf{B}$, as $\mathsf{C}=\mathsf{AB}^{t}/T$ where {the} entries of each of the matrices are real Gaussian variables with $0$ mean and variance $1$. For the uncorrelated $\mathsf{A}$ and $\mathsf{B}$, i.e. where $\mathsf{C}$ vanishes on the ensemble average, density of the singular values has been derived in Refs. \cite{Muller,Bouchaud:2007,Burda:2010,Akeman:2013}. For the correlated $\mathsf{A}$ and $\mathsf{B}$, using the CWOE approach the Pastur {equation} \cite{Pastur} has been derived in Ref. \cite{vin2013}. We consider $\overline{\mathsf{AB}^{t}}/T=\eta$, where the overbar denotes the ensemble averaging, and $\eta$ is the $N\times N$ correlation matrix which defines correlations between the rows of $\mathsf{A}$ and $\mathsf{B}$. The joint probability density of the matrix elements is described via
\begin{equation}\label{JPDAB}
P(\mathsf{A},\mathsf{B})\propto \exp
\left[-\frac{T}{2}\text{tr}\,\left\{
\left(
\begin{matrix}
\mathbf{1}_{N} & \eta
\\
\eta^{t} & \mathbf{1}_{N}
\end{matrix}
\right)
^{-1}
\left(
\begin{matrix}
\mathsf{A} \\
      \mathsf{B} 
\end{matrix}
\right)
\left(
\begin{matrix}
\mathsf{A}^{t} \, \mathsf{B}^{t} 
\end{matrix}
\right)
\right\}
\right],
\end{equation}
where matrices $\mathsf{A}$ and $\mathsf{B}$ both are of dimension $N\times T$ and $\mathbf{1}_{K}$ is an identity matrix of dimensions $K\times K$. {Often the spectral density is used in applications \cite{Biely:08,Livan:2012} and in this paper we derive the loop equations which describe the spectral density of $\mathsf{C}$ at large matrix dimension. Our derivation exploits the techniques developed in Refs. \cite{Sommers,Nowak-Nowak,Joshua-Zee,Janik-Nowak, Burda-Janik:2010} and used recently for the $\eta=0$ case in Ref. \cite{Burda:2010}}. We solve these equation and obtain a formula for the contour describing the boundary of the bulk of eigenvalues of $\mathsf{C}$. {This result is interesting from the application point of view as it states that the contour depends only on the spectrum of $\eta$ when $\eta$ is a normal matrix}, while for nonnormal $\eta$ the contour depends on its symmetric and the anti-symmetric parts. We present some examples for tridiagonal $\eta$. {In applications, however, $\eta$ may vary from system to system yet with this simple but non-trivial $\eta$ we may display details of the theory presented in this paper.} In particular, we derive the spectral density for two cases, viz., ($i$) $\eta_{jk}=c$ and ($ii$) $\eta_{jk}=c\,\delta_{jk}$. We show that in both cases the density depends only on the radial distance; however, in the latter case the contour is no longer a circle centered at origin but a shifted ellipse.

In the next section we define the nonsymmetric correlation matrices from the CWOE approach, fix notations and discuss generalities of the model. In Sec. \ref{Sec-Results}, we derive the loop equations. In section \ref{Contours}, we derive the formula for the contours defining boundaries of the bulk of complex eigenvalues of $\mathsf{C}$ for those $\eta$'s for which absolute values of the eigenvalues are bounded from above by $1$. In Sec. \ref{examples}, we compare our analytics with numerics for tridiagonal $\eta$. In Sec. \ref{densities}, we solve the loop equations {and derive the spectral density} for some especial cases, as discussed above. In Sec. \ref{Conc}, we summaries our work with conclusion.

\section{Generalities}\label{General}
The problem we address in this paper is closely related to CWOE. CWOE is an ensemble of real symmetric matrices of type $\mathcal{C}=\mathcal{WW}^{t}/T$. For CWOE, $\mathcal{W}=\xi^{1/2}\mathcal{W'}$ where $\xi$ is a real symmetric positive definite nonrandom (fixed) matrix {which takes account of the the correlations in rows of $\mathcal{W}$} and the entries of the matrix $\mathcal{W'}$ are independent real Gaussian variables with mean $0$ and variance $1$. Thus, on average we have
\begin{equation}
\overline{\mathcal{C}}=\xi.
\end{equation}
In our case the matrix $\mathcal{W}$ constitutes of two different random matrices $\mathsf{A}$ and $\mathsf{B}$, as
\begin{eqnarray}
\mathcal{W´}=
\left(
\begin{matrix}
\mathsf{A} \\
\mathsf{B} 
\end{matrix}
\right),
\end{eqnarray}
where $\mathsf{A}$ and $\mathsf{B}$ are both of dimensions $N\times T$. The ensemble correlation matrix $\xi$ we consider here is given by 
\begin{equation}\label{CWEXI}
\xi=
\left( 
\begin{matrix}
\mathbf{1}_{N} & \eta\\
\eta^{t} & \mathbf{1}_{N}
\end{matrix}
\right),
\end{equation}
where the diagonal blocks imply only self-correlations among the variables of $\mathsf{A}$ and $\mathsf{B}$. The off-diagonal $\eta$-blocks account for the correlations between $\mathsf{A}$ and $\mathsf{B}$. The correlation matrix we are interested in corresponds to the upper off-diagonal block of $\mathcal{C}$:
\begin{equation}
\mathsf{C}=\frac{1}{T}\mathsf{AB}^{t},
\end{equation}
so that $\overline{\mathsf{C}}=\eta$. Here we also define the ratio,
\begin{eqnarray}
\kappa&=&\dfrac{N}{T}.
\end{eqnarray}

Note that by construction $\mathsf{C}$ is of rank $\text{min}\{N,T\}$, therefore it will have exactly $N-T$ zero eigenvalues if $T<N$. In the following we consider a large $N$ limit, such that {$\kappa$} is finite, so that the matrix $\mathsf{C}$ will never be deterministic. Since by definition $\xi$ is a positive definite matrix, therefore $\eta$ can not be chosen arbitrarily. For instance, the positive definiteness of $\xi$ implies that the absolute values of the eigenvalues of $\eta$ are bounded from above by $1$; since the singular values have an upper bound \cite{vin2013}, Weyl's theorem \cite{Weyl} implies the same bound for the absolute values of the eigenvalues.

We are interested in the statistics of the eigenvalues of $\mathsf{C}$. For instance, the eigenvalue density, $\rho_{\mathsf{C}}(z,z^{*})$, which is defined as
\begin{equation}\label{denC}
\rho_{\mathsf{C}}(z,z^{*})=\dfrac{1}{N}\sum_{j=1}^{N}\delta(z-\lambda_{j}) \delta(z^{*}-\lambda_{j}^{*}),
\end{equation}
where $z^{*}$ stands for the complex conjugate of $z$. We use {\it overbar} to represent the ensemble averaged quantities. The ensemble averaged density can be obtained from the Green's function \cite{Nowak-Nowak}, $\overline{\mathbf{g}}_{\mathsf{C}}(z,z^{*})$, via 
\begin{equation}\label{Gauss}
\overline{\rho}_{\mathsf{C}}(z,z^{*})=\dfrac{1}{\pi}\dfrac{\partial \overline{\mathbf{g}}_{\mathsf{C}}(z,z^{*})}{\partial z^{*}},
\end{equation}
where 
\begin{equation}\label{gbarc}
\overline{\mathbf{g}}_{\mathsf{C}}(z,z^{*})=\lim_{\epsilon\rightarrow 0}
\left\langle
\overline{
\dfrac{z^{*}\mathbf{1}_{N}-\mathsf{C}^{t}}{(z^{*}\mathbf{1}_{N}-\mathsf{C}^{t})(z\mathbf{1}_{N}-\mathsf{C})+\epsilon^{2}\mathbf{1}_{N}}}
\right\rangle.
\end{equation} 
In the above definition we have introduced the spectral averaging, using the angular brackets as $\langle \mathsf{H}\rangle=\text{tr}\,\mathsf{H}/K$, for a $K\times K$ matrix $\mathsf{H}$. 

Often the above definitions are described better as an analogy of two-dimensional electrostatics \cite{Sommers, Nowak-Nowak, Joshua-Zee}. For instance, the {\it potential} in this case is $F(z,z^{*})=N^{-1}\text{ln}(\text{Det}(z^{*}-\mathsf{C}^{t})(z-\mathsf{C})+\epsilon^{2})$. The analogous {\it electric field} will then be $\overline{\mathbf{g}}_{\mathsf{C}}(z,z^{*})$. Finally, $\overline{\rho}_{\mathsf{C}}(z,z^{*})$ will be the charge density as a consequence of the two-dimensional Gauss law (\ref{Gauss}). {We refer to the Ref. \cite{Nowak-Nowak} for the details of Eqs. (\ref{Gauss}) and (\ref{gbarc}).}

In order to evaluate $ \overline{\mathbf{g}}_{\mathsf{C}}(z,z^{*})$ for large $N$, one is tempted to use the methods developed for the Hermitian matrices, e.g. the diagrammatic expansion method \cite{Brezin, Joshua-Zee, SenM} or the binary correlation method \cite{Brody81, vin2013, vp2010, ap81}. However, the large $z$-expansion, {which is used in these methods}, {contains} nonlinear combinations of $\mathsf{C}$ and $\mathsf{C}^{t}$ and these combinations make direct applications of such methods very complicated. As in \cite{Burda-Janik:2010}, to circumvent this problem we rather calculate a $2N\times 2N$ matrix, $\overline{\mathbf{G}}(z,z^{*})$, defined as
\begin{equation}\label{2dG}
\overline{\mathbf{G}}(z,z^{*})=\lim_{\epsilon\rightarrow 0}\overline{\left(
\begin{matrix}
z\mathbf{1}_{N}-\mathsf{C} & \ii \epsilon \mathbf{1}_{N}\\
\ii \epsilon\mathbf{1}_{N} &z^{*}\mathbf{1}_{N}-\mathsf{C}^{t}
\end{matrix}
\right)^{-1}},
\end{equation}
{where} $\overline{\mathbf{g}}_{\mathsf{C}}(z,z^{*})$ will be the {spectral average of the  upper diagonal block of} $ \overline{\mathbf{G}}(z,z^{*})$. {The eigenvalues of $\mathsf{C}$ are isolated poles of $\overline{\mathbf{G}}(z,z^{*})$ scattered in the complex plane. For $N\to\infty$ the eigenvalues coalesce and define a nonholomorphic region. Like the symmetric case \cite{vp2010}, an infinitesimal $\epsilon>0$ is needed to allow for analytic continuation to the nonholomorphic region.}
 
Our goal is to calculate $\overline{\mathbf{G}}(z,z^{*})$ which is still complicated due to {the non-Gaussian probability associated to the distribution of the matrix elements of $\mathsf{C}$}. To simplify the problem we use a trick of linearization, as proposed in Ref. \cite{Nowak-Nowak}, {which unfolds the problem as linear in $\mathsf{A}$ and $\mathsf{B}^{t}$ and consequently simplifies} the binary correlation method \cite{vp2010} {in a straightforward manner}. We emphasize that our results are valid in the same limit as has been addressed in Ref. \cite{Burda:2010}{, viz. when $N\to\infty$ such that $\kappa$ is finite.}

\section{The Loop Equations}\label{Sec-Results}
We begin with linearizing the problem, defining an $(N+T)\times (N+T)$ dimensional matrix $\mathsf{P}$ as
\begin{equation}\label{P}
\mathsf{P}=\dfrac{1}{\sqrt{T}}\left(
\begin{matrix}
\mathbf{0} & \mathsf{A}
\\
\mathsf{B}^{t} & \mathbf{0}
\end{matrix}
\right).
\end{equation}
{With a  note} that the nonzero eigenvalues of $\mathsf{P}^{2}$ coincide with the nonzero eigenvalues of $\mathsf{C}$ \cite{Burda-Janik:2010} and {have} a two-fold degeneracy for each, {so} we calculate $\overline{\rho}_{\mathsf{C}}(z,z^{*})$ from the spectral density, $\overline{\rho}_{\mathsf{P}}(w,w^{*})$, of $\mathsf{P}$, using
\begin{equation}\label{denP-C}
\overline{\rho}_{\mathsf{C}}(z,z^{*})=\dfrac{1}{2|z|}\overline{\rho}_{\mathsf{P}}(w(z),w^{*}(z^{*})),
\end{equation}
with $z=w^{2}$. Another very useful relation is {in} between the corresponding Green's functions:
\begin{equation}\label{MC-MP} 
z \overline{\mathbf{g}}_{\mathsf{C}}(z,z^{*})-1=\frac{T+N}{2N} \left[w \overline{\mathbf{g}}_{\mathsf{P}}(w(z),w^{*}(z^{*}))-1\right],
\end{equation}
where $\overline{\mathbf{g}}_{\mathsf{P}}(w,w^{*})$ is the Green's function of $\overline{\rho}_{\mathsf{P}}(w(z),w^{*}(z^{*}))$. {As} it will be {shown} later, we first derive $\overline{\mathbf{g}}_{\mathsf{P}}(w,w^{*})$ and consequently $\overline{\mathbf{g}}_{\mathsf{\mathsf{C}}}(z,z^{*})$ using the above relations.

{As it is stated above, in the binary correlation method it will be convenient to deal with a matrix valued Green's function}. Following the definition (\ref{2dG}), we define a $2(N+T)\times 2(N+T)$ dimensional matrix $\tilde{\mathsf{P}}$ as
\begin{equation}
\tilde{\mathsf{P}}=\left(
\begin{matrix}
\mathsf{P} & \mathbf{0}
\\
\mathbf{0} & \mathsf{P}^{t}
\end{matrix}
\right).
\end{equation}
The corresponding Green's function $\tilde{\mathsf{G}}$ can then be written as 
\begin{equation}\label{BigG}
\overline{\tilde{\mathsf{G}}}(w, w^{*})=\lim_{\epsilon\to0}\overline{\left(\tilde{\mathsf{W}}-\tilde{\mathsf{P}}
\right)^{-1}},
\end{equation}
where
\begin{equation}
\tilde{\mathsf{W}}=\left(
\begin{matrix}
w\,\mathbf{1}_{N+T} & \ii\epsilon\mathbf{1}_{N+T}
\\
\ii\epsilon\mathbf{1}_{N+T} & w^{*}\,\mathbf{1}_{N+T}
\end{matrix}
\right).
\end{equation}
{Like the relation between $\overline{\mathsf{g}}_{\mathsf{C}}(z,z^{*})$ and $\overline{\mathbf{G}}(z,z^{*})$, here as well, $\overline{\mathbf{g}}_{\mathsf{P}}(w, w^{*})$ is given by the spectral average of the $(N+T)\times(N+T)$ dimensional upper diagonal block of $\overline{\tilde{\mathsf{G}}}(w, w^{*})$.} As in \cite{Burda:2010}, it is suggestive to write $\overline{\tilde{\mathsf{G}}}$ in terms of smaller subblocks
\begin{equation}\label{Gblock}
\overline{\tilde{\mathsf{G}}}=
\left(
\begin{matrix}
\overline{\mathsf{G}}_{11} & \overline{\mathsf{G}}_{12} & \overline{\mathsf{G}}_{1\overline{1}} &\overline{\mathsf{G}}_{1\overline{2}}
\\
\overline{\mathsf{G}}_{21} & \overline{\mathsf{G}}_{22} & \overline{\mathsf{G}}_{2\overline{1}} &\overline{\mathsf{G}}_{2\overline{2}}
\\
\overline{\mathsf{G}}_{\overline{1}1} & \overline{\mathsf{G}}_{\overline{1}2} & \overline{\mathsf{G}}_{\overline{1}\,\overline{1}} &\overline{\mathsf{G}}_{\overline{1}\,\overline{2}}
\\
\overline{\mathsf{G}}_{\overline{2}1} & \overline{\mathsf{G}}_{\overline{2}2} & \overline{\mathsf{G}}_{\overline{2}\,\overline{1}} &\overline{\mathsf{G}}_{\overline{2}\,\overline{2}}
\end{matrix}
\right),
\end{equation}
where $\overline{\mathsf{G}}_{jj}$ and $\overline{\mathsf{G}}_{\overline{j}\,\overline{j}}$ are $N\times N$ and $T\times T$ respectively for the block indices $j=1$ and $2$; following the convention used in Ref. \cite{Burda:2010} we use the overbar, again but for integers, to represent $(N+T)\times(N+T)$ blocks. 

In order to perform the ensemble averaging we expand $\overline{\tilde{\mathsf{G}}}$ for large $w$: 
\begin{equation}
\label{DS-1}
\overline{\tilde{\mathsf{G}}}=\tilde{\mathsf{W}}^{-1}+\tilde{\mathsf{W}}^{-1} \overline{\tilde{\mathsf{P}}
\tilde{\mathsf{W}}^{-1}
\tilde{\mathsf{P}} 
\tilde{\mathsf{W}}^{-1}}+\dots~.
\end{equation}
As in the case of symmetric correlation matrices \cite{vp2010}, we collect only the leading order terms and avoid those resulting in ${O}(N^{-1})$ or smaller. Using the jpd (\ref{JPDAB}), we first write the following exact identities, valid for arbitrary fixed matrices $\chi_{1}$ and $\chi_{2}$: 
\begin{eqnarray}
\label{id-1}
\dfrac{1}{T}\overline{\mathsf{A} \chi_{1}\mathsf{A}^{t}\chi_{2}}&=&\langle \chi_{1}\rangle \, \chi_{2},\\
\label{id-2}
\overline{\mathsf{A} \chi_{1}\mathsf{A}\chi_{2}}&=& \chi_{2} \chi_{1}^{t},\\
\label{id-3}
\dfrac{1}{T}\overline{\mathsf{B} \chi_{1}\mathsf{B}^{t}\chi_{2}}&=&\langle \chi_{1}\rangle \, \chi_{2},\\
\label{id-4}
\overline{\mathsf{B} \chi_{1}\mathsf{B}\chi_{2}}&=& \chi_{2} \chi_{1}^{t}.
\end{eqnarray}
These identities are sufficient to obtain results for the $\eta=0$ case. However, for $\eta\ne0$, we shall also use 
\begin{eqnarray}
\label{id-5}
\dfrac{1}{T}\overline{\mathsf{A} \chi_{1}\mathsf{B}^{t}\chi_{2}}&=&\langle \chi_{1}\rangle \, \eta \chi_{2},\\
\label{id-6}
\dfrac{1}{T}\overline{\mathsf{A}^{t} \chi_{1}\mathsf{B}\chi_{2}}&=&\kappa\,\langle \eta^{t}\chi_{1}\rangle \, \chi_{2},\\
\label{id-7}
\dfrac{1}{T}\overline{\mathsf{B} \chi_{1}\mathsf{A}^{t}\chi_{2}}&=&\langle \chi_{1}\rangle \, \eta^{t}\chi_{2},\\
\label{id-8}
\dfrac{1}{T}\overline{\mathsf{B}^{t} \chi_{1}\mathsf{A}\chi_{2}}&=&\kappa\,\langle \eta\chi_{1}\rangle \, \chi_{2}.
\end{eqnarray}
In what follows, we ignore $O(N^{-1})$ terms in the ensemble averaging while keeping only the leading order terms. {We use} the identities (\ref{id-1}, \ref{id-3}) and (\ref{id-5}-\ref{id-8}), i.e., avoiding terms resulting from the binary associations of $\mathsf{A}$ with $\mathsf{A}$ and $\mathsf{B}$ with $\mathsf{B}$. Terms resulting from the binary association of $\mathsf{A}$ with $\mathsf{B}$ are $0$, thus will also be ignored. We {then} obtain
\begin{equation}\label{DS-2}
\overline{\tilde{\mathsf{G}}}= 
\left(\tilde{\mathsf{W}}-\Sigma
\right)^{-1},
\end{equation}
where
\begin{equation}
\Sigma= 
\left(
\begin{matrix}
\langle\overline{\mathsf{G}}_{22}\rangle\eta & \mathbf{0} & \langle\overline{\mathsf{G}}_{2\overline{2}}\rangle \mathbf{1}_{N} & \mathbf{0}\\
\mathbf{0} & \kappa \langle\eta\overline{\mathsf{G}}_{11}\rangle \mathbf{1}_{T} & \mathbf{0} & \kappa \langle\overline{\mathsf{G}}_{1\overline{1}}\rangle \mathbf{1}_{T}\\
\langle\overline{\mathsf{G}}_{\overline{2}2}\rangle\mathbf{1}_{N} & \mathbf{0} & \langle\overline{\mathsf{G}}_{\overline{2}\,\overline{2}}\rangle\eta^{t}& \mathbf{0} \\
\mathbf{0} & \kappa \langle\overline{\mathsf{G}}_{\overline{1}1}\rangle \mathbf{1}_{T} & \mathbf{0} & \kappa \langle\overline{\eta^{t}\mathsf{G}}_{\overline{1}\,\overline{1}}\rangle\mathbf{1}_{T}
\end{matrix}
\right).
\end{equation}
{Notice} that avoiding the $O(N^{-1})$ or smaller contributions {yields} zeros in the rectangular blocks. Moreover, {like the symmetric case \cite{vrt2013}}, in the derivation of Eq.(\ref{DS-2}) we have also avoided the binary associations across the spectral averages as those will also produce terms of $O(N^{-1})$. Finally, calculating the inverse in Eq. (\ref{DS-2}), we obtain
\begin{equation}\label{DSF}
\overline{\tilde{\mathsf{G}}}= 
\left(
\begin{matrix}
 \overline{\Gamma}^{(1)}_{1} \overline{\mathsf{d}}_{1}& \mathbf{0}&  \overline{\Gamma}^{(1)}_{1} \langle\overline{G}_{2\,\overline{2}}\rangle& \mathbf{0}\\
 \mathbf{0} & \overline{\Gamma}^{(2)}_{1} \overline{\mathsf{d}}_{2} &\mathbf{0} & \kappa \overline{\Gamma}^{(2)}_{1} \langle\overline{G}_{1\,\overline{1}}\rangle\\
\overline{\Gamma}^{(1)}_{2} \langle\overline{G}_{\overline{2}\,2}\rangle &\mathbf{0} &  \overline{\Gamma}^{(1)}_{2} \overline{\mathsf{a}_{1}}& \mathbf{0}\\
\mathbf{0} & \kappa \overline{\Gamma}^{(2)}_{2} \langle\overline{G}_{\overline{1}\,1}\rangle& \mathbf{0} & \overline{\Gamma}^{(2)}_{2} \overline{\mathsf{a}}_{2}
\end{matrix}
\right).
\end{equation}
Here, 
\begin{eqnarray}\label{gam11}
\overline{\Gamma}^{(1)}_{1}&=&\left[\overline{\mathsf{d}}_{1}\overline{\mathsf{a}}_{1}-\overline{g}_{2\overline{2}}\overline{g}_{\overline{2}2}\mathbf{1}_{N}\right]^{-1},\\
\label{gam12}
\overline{\Gamma}^{(1)}_{2}&=&\left[\overline{\mathsf{a}}_{1}\overline{\mathsf{d}}_{1}-\overline{g}_{2\overline{2}}\overline{g}_{\overline{2}2}\mathbf{1}_{N}\right]^{-1},\\
\label{gam21}
\overline{\Gamma}^{(2)}_{1}&=&\left[\overline{\mathsf{d}}_{2}\overline{\mathsf{a}}_{2}-\kappa^{2}\overline{g}_{1\overline{1}}\overline{g}_{\overline{1}1}\mathbf{1}_{T}\right]^{-1},\\
\label{gam22}
\overline{\Gamma}^{(2)}_{2}&=&\left[\overline{\mathsf{a}}_{2}\overline{\mathsf{d}}_{2}-\kappa^{2}\overline{g}_{1\overline{1}}\overline{g}_{\overline{1}1}\mathbf{1}_{T}\right]^{-1},
\end{eqnarray}
and
\begin{eqnarray}\label{a1d1}
&&\overline{\mathsf{a}}_{1}= w\mathbf{1}_{N}-\overline{g}_{22}\eta,~\overline{\mathsf{d}}_{1}= w^{*}\mathbf{1}_{N}-\overline{g}_{\overline{2}\,\overline{2}}\eta^{t},\\
\label{a2d2}
&&\overline{\mathsf{a}}_{2}= (w-\kappa \overline{g}_{11,\eta})\mathbf{1}_{T},~\overline{\mathsf{d}}_{2}= (w^{*}-\kappa\overline{g}_{\overline{1}\,\overline{1},\eta^{t}})\mathbf{1}_{T}.
\end{eqnarray}
{Some details of the derivation of the above equation are given in Appendix \ref{apInv}}.

In the above result we have used rather a more general spectral-averaging, viz.,
\begin{eqnarray}\label{gLjj}
&&\overline{g}_{jj,\mathsf{L}}=\langle \mathsf{L} \overline{\mathsf{G}}_{jj}\rangle,~\overline{g}_{\overline{j}\,\overline{j},\mathsf{L}}=\langle \mathsf{L}\overline{\mathsf{G}}_{\overline{j}\,\overline{j}}\rangle,
\end{eqnarray}
where $\mathsf{L}$ is $N\times N$ {or} $T\times T$, respectively for $j=1$ {or} $2$. For instance, in Eq. (\ref{a2d2}), we have considered $\mathsf{L}=\eta$ and $\eta^{t}$, respectively in the first and the second equality. {Similarly we have used $\overline{g}_{j\,\overline{j}}=\langle\overline{\mathsf{G}}_{j\,\overline{j}}\rangle$ and $\overline{g}_{\overline{j}\,j}=\langle\overline{\mathsf{G}}_{\overline{j}\,j}\rangle$ for the spectral averages of the off-diagonal blocks with their respective dimensions.}

Next, on equating the right-hand sides of Eqs. (\ref{Gblock}) and (\ref{DSF}), we derive a set of coupled equations. Comparing spectral averages of the diagonal blocks {of these equations} we {obtain}
\begin{eqnarray}\label{res1-gjj}
\overline{g}_{11}&=&\langle \overline{\Gamma}^{(1)}_{1} \overline{\mathsf{d}}_{1}\rangle,~\overline{g}_{22}=\langle \overline{\Gamma}^{(2)}_{1}  \overline{\mathsf{d}}_{2}\rangle,\\
\label{res2-gjj}
\overline{g}_{\overline{1}\,\overline{1}}&=&\langle \overline{\Gamma}^{(1)}_{2} \overline{ \mathsf{a}}_{1}\rangle,~
\overline{g}_{\overline{2}\,\overline{2}}=\langle \overline{\Gamma}^{(2)}_{2}  \overline{\mathsf{a}}_{2}\rangle.
\end{eqnarray}
Similarly, averaging over the off-diagonal blocks gives
\begin{eqnarray}\label{res-g11b}
\overline{g}_{1\overline{1}}&=&\overline{g}_{2\overline{2}}\langle \overline{\Gamma}^{(1)}_{1} \rangle,~
\overline{g}_{\overline{1}1}=\overline{g}_{\overline{2}2}\langle \overline{\Gamma}^{(1)}_{2} \rangle,\\
\label{res-g22b}
\overline{g}_{2\overline{2}}&=&\kappa\overline{g}_{1\overline{1}}\langle \overline{\Gamma}^{(2)}_{1} \rangle,~
\overline{g}_{\overline{2}2}=\kappa\overline{g}_{\overline{1}1}\langle \overline{\Gamma}^{(2)}_{2} \rangle.
\end{eqnarray}
All together the Eqs. (\ref{res1-gjj},\ref{res2-gjj},\ref{res-g11b},\ref{res-g22b}) are the loop equations for $\mathsf{P}$. Eliminating $\overline{g}_{1\overline{1}}$ and $\overline{g}_{\overline{1}1}$ from Eqs. (\ref{res-g11b},\ref{res-g22b}) we get an important identity:
\begin{equation}\label{Id-gam1gam2}
\langle \overline{\Gamma}^{(1)}_{j} \rangle\, \langle \overline{\Gamma}^{(2)}_{j} \rangle=\dfrac{1}{\kappa},
\end{equation}
for $j=1,2$. The Green's function, which describes the spectral density for $\mathsf{P}$, is {formally} given by \cite{Burda:2010}
\begin{equation}\label{F-Result-gp}
\overline{\mathbf{g}}_{\mathsf{P}}(w,w^{*})= \frac{N\, \overline{g}_{11}(w, w^{*})+ T\,\overline{g}_{22}(w, w^{*})}{N+T}.
\end{equation}

As in the previous cases \cite{Nowak-Nowak,Burda:2010}, here we {obtain} two solutions for $\overline{g}_{jj}$. For the trivial one, $\overline{g}_{\overline{j}j}$ and $\overline{g}_{j\overline{j}}$ are $0$. This solution corresponds to the holomorphic region as for large $w$ the Green's function behaves as $1/w$. The other {solution} is a nontrivial, giving $\overline{\mathbf{g}}_{\mathsf{P}}(w,w^{*})$, and consequently $\overline{\mathbf{g}}_{\mathsf{\mathsf{C}}}(z,z^{*})$ for which the Gauss law (\ref{Gauss}) gives the density we desire in the nonholomorphic region. For infinitely large matrices, however, the density has a sharp cut-off at the boundary of holomorphic and the nonholomorphic regions.

\section{The Domain of Eigenvalues}\label{Contours}

For a general $\eta$, {an} analytic solution in the nonholomorphic region is difficult to obtain. However, the situation is simpler at the boundary of the holomorphic-nonholomorphic region. As in Refs. \cite{Nowak-Nowak,Janik-Nowak,Burda-Janik:2010, Burda:2010}, we match the solution of nonholomorphic and holomorphic regions and derive {a} formula for the contours defining such boundaries. {For this purpose we solve the loop equations in the nonholomorphic region and then use the solutions corresponding the holomorphic region, viz. $\overline{g}_{j\overline{j}}=\overline{g}_{\overline{j}j}=0$. At first we note that with $\overline{g}_{j\overline{j}}=\overline{g}_{\overline{j}j}=0$, Eqs. (\ref{gam11}-\ref{gam22}) simplify to }
\begin{eqnarray}
\langle \overline{\Gamma}^{(1)}_{1}\rangle&=&\langle\left[\overline{\mathsf{d}}_{1}\overline{\mathsf{a}}_{1}\right]^{-1}\rangle=\langle \overline{\Gamma}^{(1)}_{2}\rangle\equiv \overline{\gamma}_{1},\\
\langle \overline{\Gamma}^{(2)}_{1}\rangle&=&\langle\left[\overline{\mathsf{d}}_{2}\overline{\mathsf{a}}_{2}\right]^{-1}\rangle=\langle \overline{\Gamma}^{(2)}_{2}\rangle\equiv \overline{\gamma}_{2}.
\end{eqnarray}

{For $\overline{g}_{11,\eta}$, we use the definition (\ref{gLjj}) in the first equality of Eq. (\ref{res1-gjj}) and simplify this equation by setting $\overline{g}_{j\overline{j}}=\overline{g}_{\overline{j}j}=0$.  Next, we write $\overline{g}_{22}$ in terms of $\overline{g}_{11,\eta}$ by exploiting the second equality of Eq. (\ref{res1-gjj}) with $\overline{g}_{j\overline{j}}=\overline{g}_{\overline{j}j}=0$. Finally we set $\epsilon=0$.} 
This method leads to the self-consistent equation:
\begin{equation}
\overline{g}_{11,\eta}=(w-\kappa \overline{g}_{11,\eta})
\left\langle
\eta\left(w(w-\kappa\overline{g}_{11,\eta})-\eta
\right)^{-1}
\right\rangle.
\end{equation}
Denoting $w(w-\kappa\overline{g}_{11,\eta})=\Psi$, one can cast this equation into
\begin{equation}
\label{z-Psi1}
z=\Psi+\kappa\Psi\left\langle
\eta\left(\Psi-\eta
\right)^{-1}
\right\rangle,
\end{equation}
where we have replaced $w^{2}$ by $z$ to transform the equation for {the eigenvalues of} $\mathsf{C}$. Similarly, using Eq. (\ref{res2-gjj}) for $\overline{g}_{\overline{1}\,\overline{1},\eta^{t}}$, we obtain
\begin{equation}
\label{z-Psi2}
z^{*}=\Psi^{*}+\kappa\Psi^{*}\left\langle
\eta^{t}\left(\Psi^{*}-\eta^{t}
\right)^{-1}
\right\rangle,
\end{equation}
{where we have used $w^{*}(w^{*}-\kappa\overline{g}_{\overline{1}\,\overline{1},\eta^{t}})=\Psi^{*}$}.

Next from the identity (\ref{Id-gam1gam2}) we obtain, {a relation $\overline{\gamma}_{1}\overline{\gamma}_{2}=\kappa^{-1}$ and thus,} an equation for $\Psi$:
\begin{equation}
\label{Contour}
\langle[(\Psi-\eta)(\Psi^{*}-\eta^{t})]^{-1}\rangle=\frac{1}{\kappa}.
\end{equation}
{Notice that the identity (\ref{Id-gam1gam2}) is valid only in the nonholomorphic region and to obtain the above equation we have used the same method employed in the derivation of Eqs. (\ref{z-Psi1}) and (\ref{z-Psi2}).} 

Equation (\ref{z-Psi1}), describes the contour enclosing the bulk of the eigenvalues in terms of the solution $\Psi$ of the Eq. (\ref{Contour}). Let $\eta$ be commuting with $\eta^{t}$, meaning $\eta$ is a normal matrix. Then Eq. (\ref{Contour}) has eigenvalue expansion and thus can be simplified in terms of the eigenvalues of $\eta$; see Sec. \ref{examples} for examples. On the other hand if $\eta$ is nonnormal then Eq. (\ref{Contour}) does not have further simplifications. The underlying remark for the latter case is that the solution of Eq. (\ref{Contour}) depends on $\eta_{S}$ and $\eta_{A}$, where $\eta_{S}=(\eta+\eta^{t})/2$ and $\eta_{A}=(\eta-\eta^{t})/2$ are respectively the symmetric and anti-symmetric parts of $\eta$. Therefore the contour (\ref{z-Psi1}) as well depends on the symmetric and the anti-symmetric parts of $\eta$. 

\section{Examples and Numerics}\label{examples}
As far as numerics are concerned, we must emphasize that one should be careful while diagonalizing $\mathsf{C}$ for large $N$, as such diagonalizations need extended precisions to obtain reasonably correct results \cite{Wilkinson}. Indeed, a very high precision is used for all numerical examples discussed below. It is worth to point out that our theory stands for a very large $N$ where the diagonalization is really expensive. Therefore we expect deviations from the theory due to the finite size of $\mathsf{C}$. {In the numerical examples, we consider $N=512$ and $T=2N$ except for Fig. \ref{Sval} where we consider $N=1024$.}

\begin{figure}
        \centering
               \includegraphics [width=0.5\textwidth]{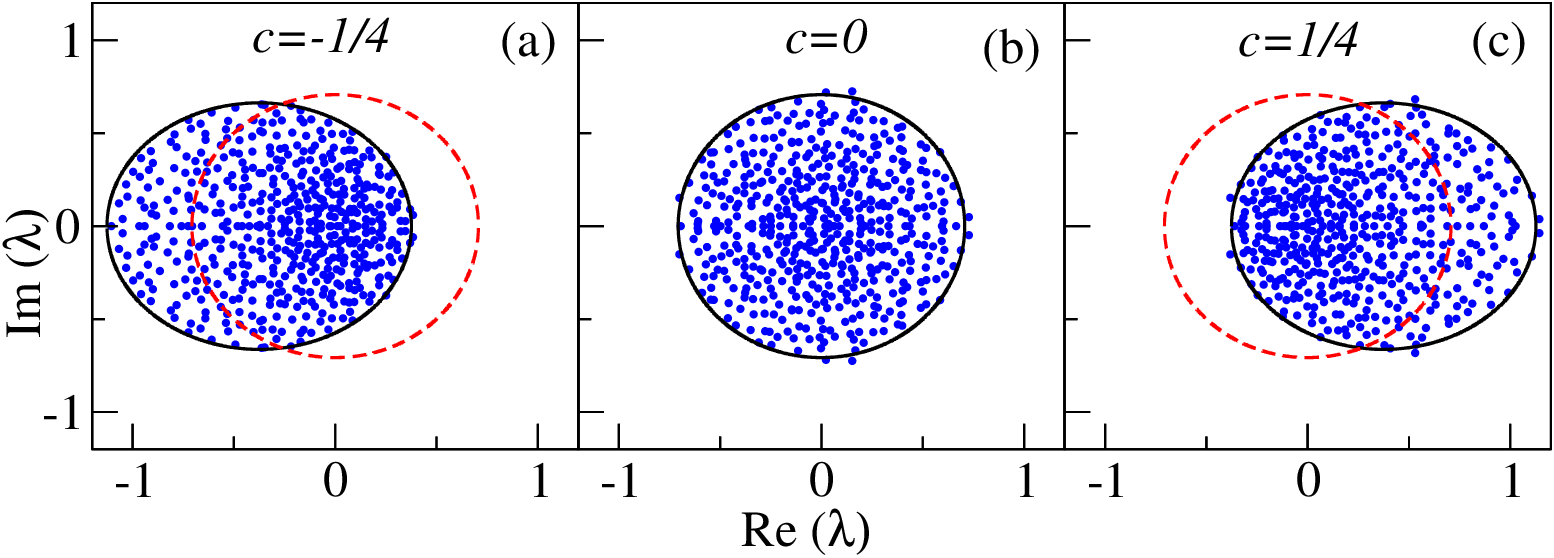}
                \caption{(Color online) Scatter plot for the eigenvalues for the case ($i$), where $\eta$ is diagonal with the same entries $c$. Here data (eigenvalues), obtained from the diagonalization of the corresponding matrix $\mathsf{C}$, are represented by dots and theory (\ref{ellipse}) is represented by the solid lines. Dashed {\it red} lines represent the theory for $c=0$. In this figure the matrix dimension $N=512$ and we show only $1$ realization.}
\label{dif-d}
\end{figure}

Below, we consider $\eta$ to be a tridiagonal matrix, $\eta_{jk}=c\,\delta_{jk}+p\,\delta_{j,k+1}+q\,\delta_{j+1,k}$, where $c,\, p$ and $q$ are real. The eigenvalues, $\lambda^{(\eta)}_{j}$, of $\eta$, are given by \cite{Kulk}
\begin{equation}
\lambda^{(\eta)}_{j}=c+2\sqrt{pq}\cos(\frac{j\pi}{N+1}),
\end{equation}
for $1\le j\le N$. To ensure the positive definiteness of $\xi$, we choose $c$ and $p,q$ such that $|\lambda^{(\eta)}_{j}|<1$. {This choice of $\eta $ is simple but capable of displaying subtleties of the theory as discussed below.} 

Let $p=q$. Then using $\lambda^{(\eta)}_{j}$ in Eq. (\ref{z-Psi1}), we can write
\begin{eqnarray}
z&=&\Psi (1-\kappa)+\kappa\Psi^{2}\sum_{j=1}^{N}\frac{1}{N\left(\Psi-c-2c_{0}\cos\left(\frac{j\pi}{N+1}\right)\right)},\nonumber\\
\end{eqnarray}
where $c_{0}=|\sqrt{pq}|$. For large $N$, the summation may be replaced by an integral. Solving this integral by using the technique of contour integration, we obtain
\begin{eqnarray}
\label{trigz-Psi}
z&=&\Psi (1-\kappa)+\frac{\kappa \Psi^{2}}{\sqrt{(\Psi-c)^{2}-4c_{0}^{2}}}.
\end{eqnarray}
Similarly, from (\ref{Contour}) we get
\begin{equation}\label{trig-Cont}
\frac{1}{k}+\frac{2 r s \cos(\theta-\nu)}{8c_{0}^{2}r^{2}\cos(2\theta)-[16c_{0}^{4}+r^{4}-s^{2}(4c_{0}^{2}-r^{2})]}=0,
\end{equation}
where 
\begin{equation}
\Psi-c=r\exp(\ii \theta),~\text{and}~\sqrt{r^{2}\exp(2\ii\theta)-4c_{0}^{2}}=s\exp(\ii\nu).
\end{equation}

We consider five cases for this tridiagonal matrix, viz., ($i$) $c\ne0$ and $p=q=0$, ($ii$) $p=q\ne 0$, ($iii$) $c=0$, $p=-q$, ($iv$) $c\ne0$, $p=-q$, and ($v$) $c\ne0$, $p\ne q$. Note that for the cases ($i,ii$) $\eta$ is symmetric and for the case ($iii$) it is anti-symmetric. In the case ($iv$) $\eta$ is nonsymmetric but commutes with $\eta^{t}$ while in ($v$) $\eta$ is nonsymmetric and does not commute with $\eta^{t}$. In the last case, Eq. (\ref{Contour}) can not be simplified to the eigenvalues and consequently Eq. (\ref{trig-Cont}) does not apply. However, for the remaining cases the latter {may} be used.

\begin{figure}[!t]
        \centering
               \includegraphics [width=0.45\textwidth]{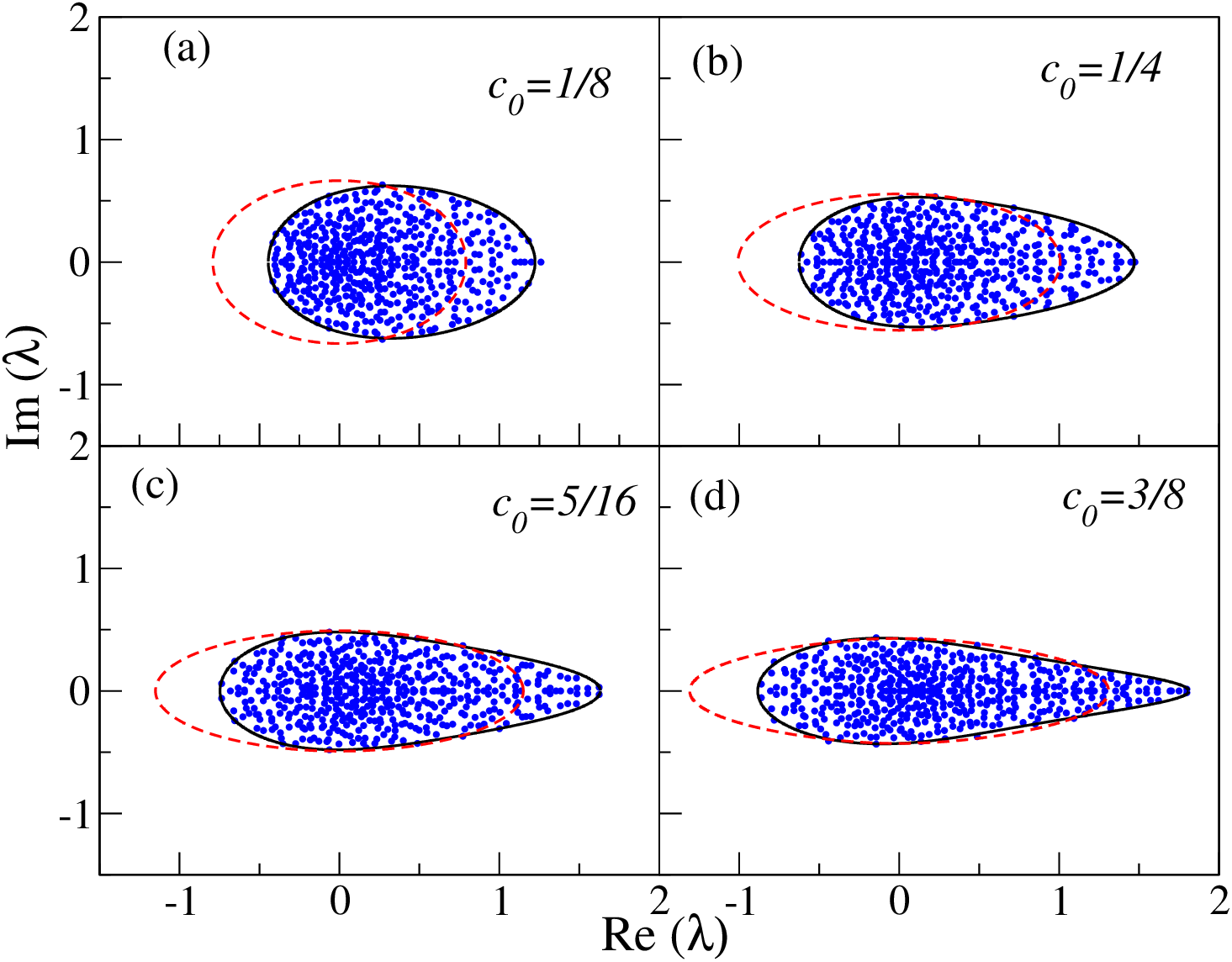}
                \caption{(Color online) Same as Fig. \ref{dif-d} for the case ($ii$) which is described in the text. The theory is compared with the numerical results for $c=1/4$ and $c_{0}=1/8,\,1/4,\,5/16$ and $3/8$ respectively in (a), (b), (c) and (d). Dashed {\it red} lines represent theory for the same $c_{0}$ values where $c=0$.}
\label{trigSym}
\end{figure}

For the case ($i$) Eq. (\ref{trig-Cont}) yields a circle, i.e., $|\Psi-c|^{2}=\kappa$. Using the solution of (\ref{trigz-Psi}) for this circle, we get
\begin{equation}\label{eq1t0}
\left|
z-c(1+\kappa)+\sqrt{[z-c(1+\kappa)]^{2}-4c^{2}\kappa}
\right|^{2}=4\kappa.
\end{equation}
Let $z=x+\ii y$. Then an ellipse can be observed from the above equation by shifting the $x$-axis, defining $Z=X+\ii Y$, where $X=x-c(1+\kappa)$ and $Y=y$, and then rescaling $X$ and $Y$ as $\phi=u+\ii v$ where $u=X/\sqrt{\kappa}(1+c^{2})$ and $v=Y/\sqrt{\kappa}(1-c^{2})$. Since Eq. (\ref{eq1t0}) is satisfied for $|\phi|=1$, {the} solution for $z$ satisfying equation (\ref{eq1t0}) is an ellipse:
\begin{equation}\label{ellipse}
\frac{\left(x-c(1+\kappa)\right)^{2}}{\kappa(1+c^{2})^{2}}+\frac{y^{2}}{\kappa(1-c^{2})^{2}}=1.
\end{equation}
In Sec. \ref{densities}, we calculate explicitly the spectral density for this case.

In Fig. \ref{dif-d} we compare these results for different $c$ with eigenvalues obtained from diagonalizing $N=512$ dimensional matrices. In this figure we show the results for $c=-1/4, 0$ and $1/4$ respectively in (a), (b) and (c). For $c=0$ the eigenvalues are enclosed by a circle of radius $\sqrt{\kappa}$ while for nonzero $c$ these are enclosed by shifted ellipses as predicted by our theory (\ref{ellipse}). 

{We observe} that for given $c_{0}$, (\ref{trig-Cont}) is the same for the cases ($ii$), ($iii$) and ($iv$), and only (\ref{trigz-Psi}) may differ. For instance, in the case ($ii$) Eq. (\ref{trigz-Psi}) remains the same but for the remaining two cases we have
\begin{eqnarray}\label{adtrigz-Psi}
z=\Psi'(1-\kappa)+\frac{\kappa\Psi'^{2}}{\ii\sqrt{\Psi^{2}-4c_{0}^{2}}},
\end{eqnarray}
where
$\Psi'=\ii\Psi+c$. In the case ($iii$) we have $\Psi'=\ii \Psi$ which gives the same contour representing its symmetric counterpart but with a change in phase by $\pi/2$.
\begin{figure}[!t]
        \centering
               \includegraphics [width=0.45\textwidth]{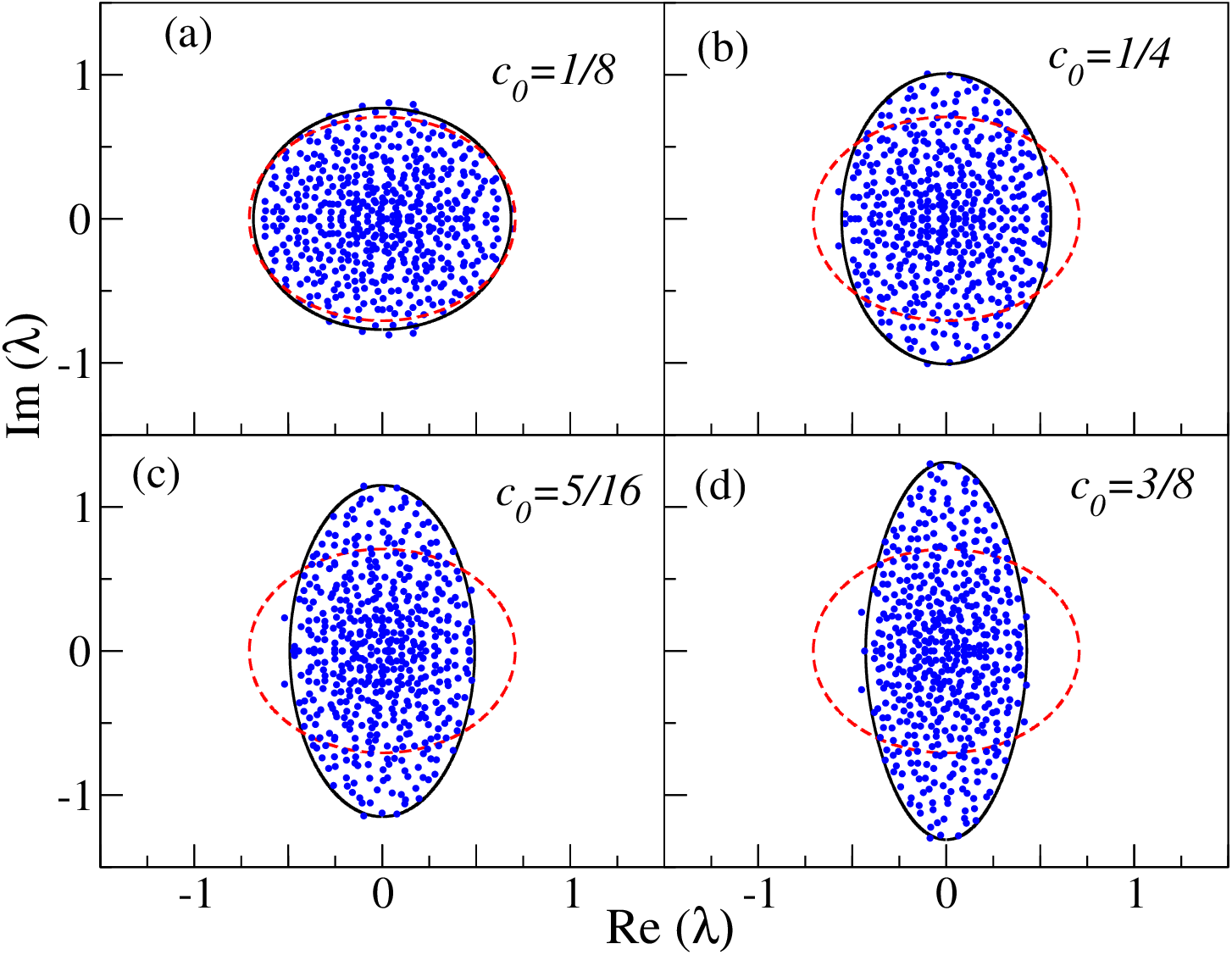}
                \caption{(Color online) Repeated on the same pattern of Fig. \ref{trigSym} for the case case ($iii$) where dashed {\it red} lines represent circles of radius $\sqrt{\kappa}$ as predicted by the theory for $c=0$.}
\label{Asymtrig}
\end{figure}
\begin{figure}
        \centering
               \includegraphics [width=0.45\textwidth]{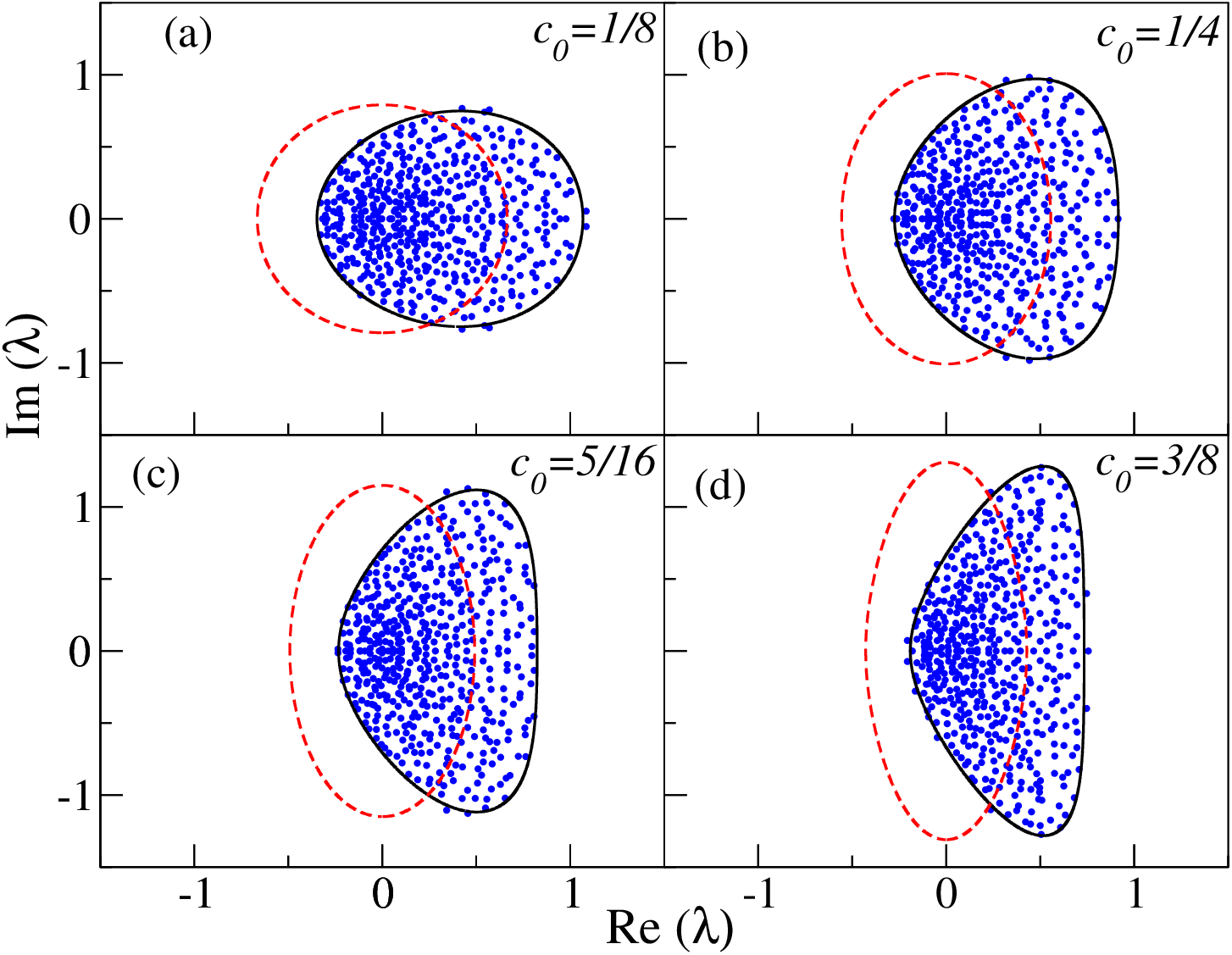}
                \caption{(Color online)  Repeated on the same pattern of Fig. \ref{trigSym} for the case case ($iv$) where dashed {\it red} lines represent the theory for $c=0$.} 
\label{Pd-anti}
\end{figure}

In Fig. \ref{trigSym}, we compare our theory with numerics for $N=512$ where $c=1/4$ and $c_{0}$ is varied from $1/4$ to $3/8$. As shown in the figure, the contours are symmetric along the real axis but not along the imaginary axis. This asymmetry arises from {the} nonzero choice of $c$. For comparison, we also show {the result} for $c=0$ where indeed the contour is symmetric along the imaginary axis.

In Fig. \ref{Asymtrig}, we vary $c_{0}$ from $1/8$ to $3/8$ to  compare our theory with numerics for the case ($iii$). {As} it is evident from the theory, in this case the contours are symmetric along both axes. For the nonsymmetric but commuting $\eta$ and $\eta^{t}$, i.e., case ($iv$), we show the results in Fig. \ref{Pd-anti} where $c_{0}$ is varied in the same range as in Fig. \ref{Asymtrig}. Here we have considered $c=1/4$, where the theoretical results are shown by solid black lines. For comparison we also plot the predictions for $c=0$ using dashed red lines. As shown in the figure, our theory gives reasonable account of the eigenvalue domain.

\begin{figure}
        \centering
               \includegraphics [width=0.45\textwidth]{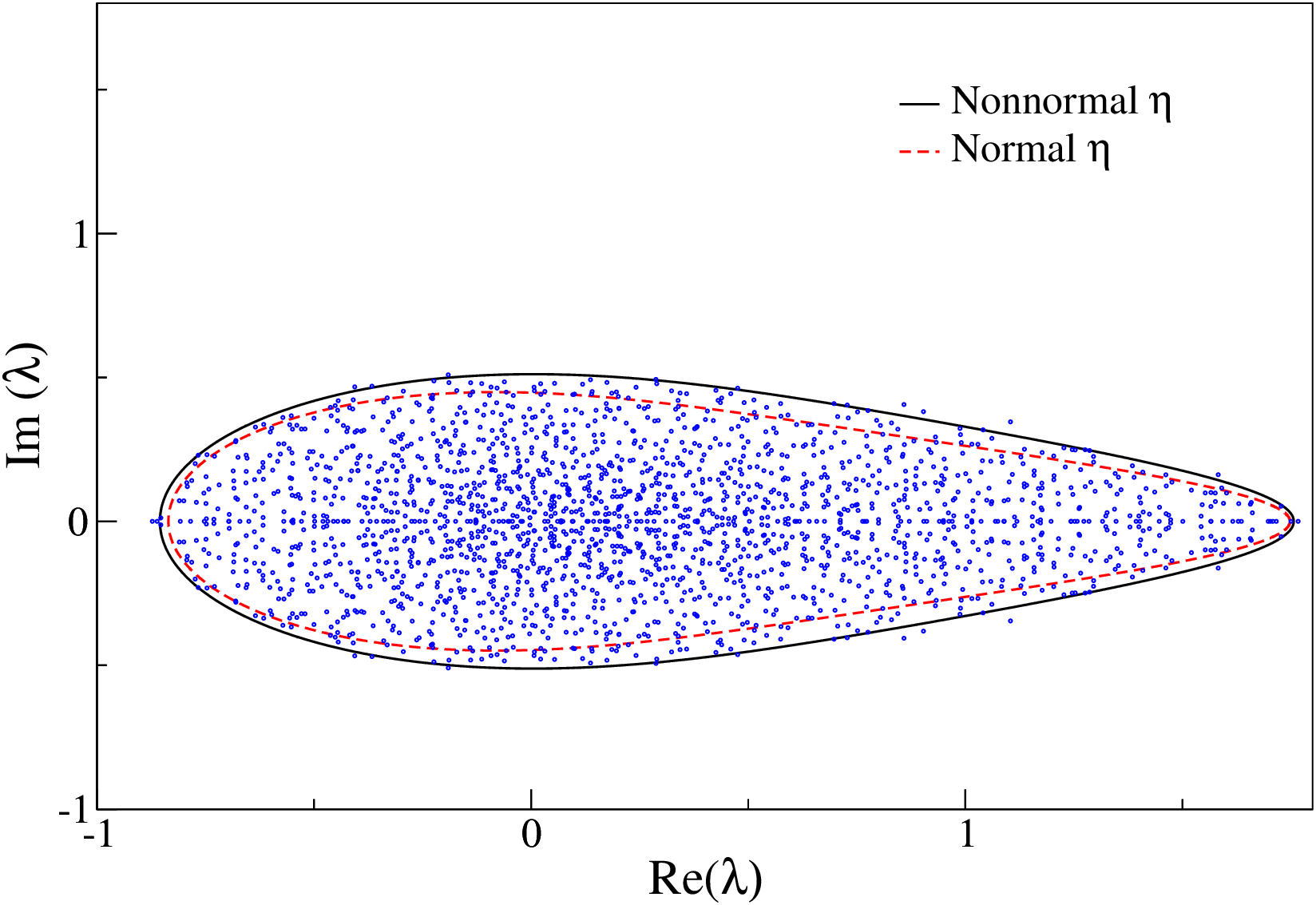}
                \caption{(Color online) 
Scatter plot for the eigenvalues of $\mathsf{C}$ corresponding non-commuting $\eta$, described as the case ($v$) in the text. Data (eigenvalues) obtained from the diagonalization of $\mathsf{C}$ are represented by dots and theory (\ref{z-Psi1}) is represented by the solid lines. Black lines represent contours for the nonsymmetric $\eta$ and red lines represent contours for the symmetric $\eta$ with the same spectrum. Results for the nonnormal case are shown for $c=1/4$, $p=1/4$ and $q=1/2$, and where we also plot theory for the normal (symmetric) case where $c=1/4$, $p=q=\sqrt{1/8}$. In this figure the matrix dimension $N=512$ and we show $4$ independent realizations.}
\label{nonsym}
\end{figure}

Finally, for the case ($v$), we consider $c=1/4$, $p=1/4$ and $q=1/2$ and $N=512$. Since in this case $\eta$ does not commute with $\eta^{t}$, Eq. (\ref{Contour}) has to be solved numerically. In Fig. \ref{nonsym}, we compare this solution with the data obtained from the matrix diagonalization. For the numerical solution of Eq. (\ref{Contour}), we consider $N=512$. As shown in the figure, our theory closely predicts the boundary. For comparison we also plot the contour which describes the boundary for the symmetric (normal) $\eta$ which has the same spectrum. This figure confirms that the boundaries are sensitive to the non-normality in the correlation matrix $\eta$ as opposed to the normal $\eta$ cases where it depends only {on} the spectrum of $\eta$. Also the singular values are sensitive to such non-normality. In Fig. \ref{Sval} we compare densities of the square of the singular values of $\mathsf{C}$ for the same normal and the nonnormal $\eta$'s. This figure confirms that the non-normality can also be observed in the singular values of $\mathsf{C}$ but for small eigenvalues or in the tail.

\section{The Density of Eigenvalues: some especial cases}\label{densities}
Let $\eta=0$. Then from Eqs. (\ref{a1d1}-\ref{a2d2}), we have $\overline{\mathsf{a}}_{1}=w\mathbf{1}_{N}$, $\overline{\mathsf{a}}_{2}=w\mathbf{1}_{T}$, $\overline{\mathsf{d}}_{1}=w^{*}\mathbf{1}_{N}$ and $\overline{\mathsf{d}}_{2}=w^{*}\mathbf{1}_{T}$. These will lead to the same set of equations as obtained in Ref. \cite{Burda:2010} for the uncorrelated matrices. On the other hand, for $\eta\ne0$, these equations are very complicated. Analytically, {one may attempt to find a linear relation} between $\langle \overline{\Gamma}^{(1)}_{j} \rangle$ and $\langle \overline{\Gamma}^{(2)}_{j} \rangle$ {because then, together with the identity (\ref{Id-gam1gam2}), this relation forms a quadratic equation for the $\langle \overline{\Gamma}^{(1)}_{j} \rangle$'s, or equivalently for the $\langle \overline{\Gamma}^{(2)}_{j} \rangle$'s. However, the $\langle \overline{\Gamma}^{(1)}_{j} \rangle$'s or the $\langle \overline{\Gamma}^{(2)}_{j} \rangle$'s are still not sufficient to obtain $\overline{g}_{jj}$, that we need to obtain $\overline{\mathsf{g}}_{\mathsf{P}}$; see Eqs. (\ref{res1-gjj}) and (\ref{F-Result-gp})}. Below we consider some especial cases where a linear relation holds {between the $\langle \overline{\Gamma}^{(1)}_{j} \rangle$'s and $\langle \overline{\Gamma}^{(2)}_{j} \rangle$'s and in these cases the $\overline{g}_{jj}$'s can be obtained from $\langle \overline{\Gamma}^{(1)}_{1} \rangle$ and $ \langle \overline{\Gamma}^{(2)}_{1} \rangle$.}

In the first case we consider $\eta_{jk}=c$ where $0\le c\le1/N$. This choice defines the equal-cross-correlation matrix model. In this case where $\eta$ is rank-$1$, it has only one nonzero eigenvalue $Nc$. On the other hand for the bulk of the spectrum \cite{bulk}, we have $\overline{\mathsf{a}}_{1}=w\mathbf{1}_{N}$, $\overline{\mathsf{a}}_{2}=w\mathbf{1}_{T}$, $\overline{\mathsf{d}}_{1}=w^{*}\mathbf{1}_{N}$ and $\overline{\mathsf{d}}_{2}=w^{*}\mathbf{1}_{T}$. These will lead to the same set of equations as one gets for the uncorrelated case. Therefore, for the bulk we know the result. For instance, using Eq. (\ref{ellipse}) we know that the contour enclosing the bulk is a circle of radius $\sqrt{\kappa}$. However, the spectrum may have one eigenvalue lying off the circle. If we avoid the bulk effect, then the ensemble averaged position of this eigenvalue is $\sim Nc$ provided $Nc>\sqrt{\kappa}$. Such observations have been very useful in the analysis of symmetric correlation matrices \cite{Finance1,Finance2,Finance3,Finance4} and also for density matrices \cite{vmarko}. On the other hand, there is an upper bound $Nc<1$ due to the positive definiteness of $\xi$ which consequently implies that there will be no eigenvalue separation from the bulk for $\kappa>1$.

\begin{figure}
        \centering
               \includegraphics [width=0.45\textwidth]{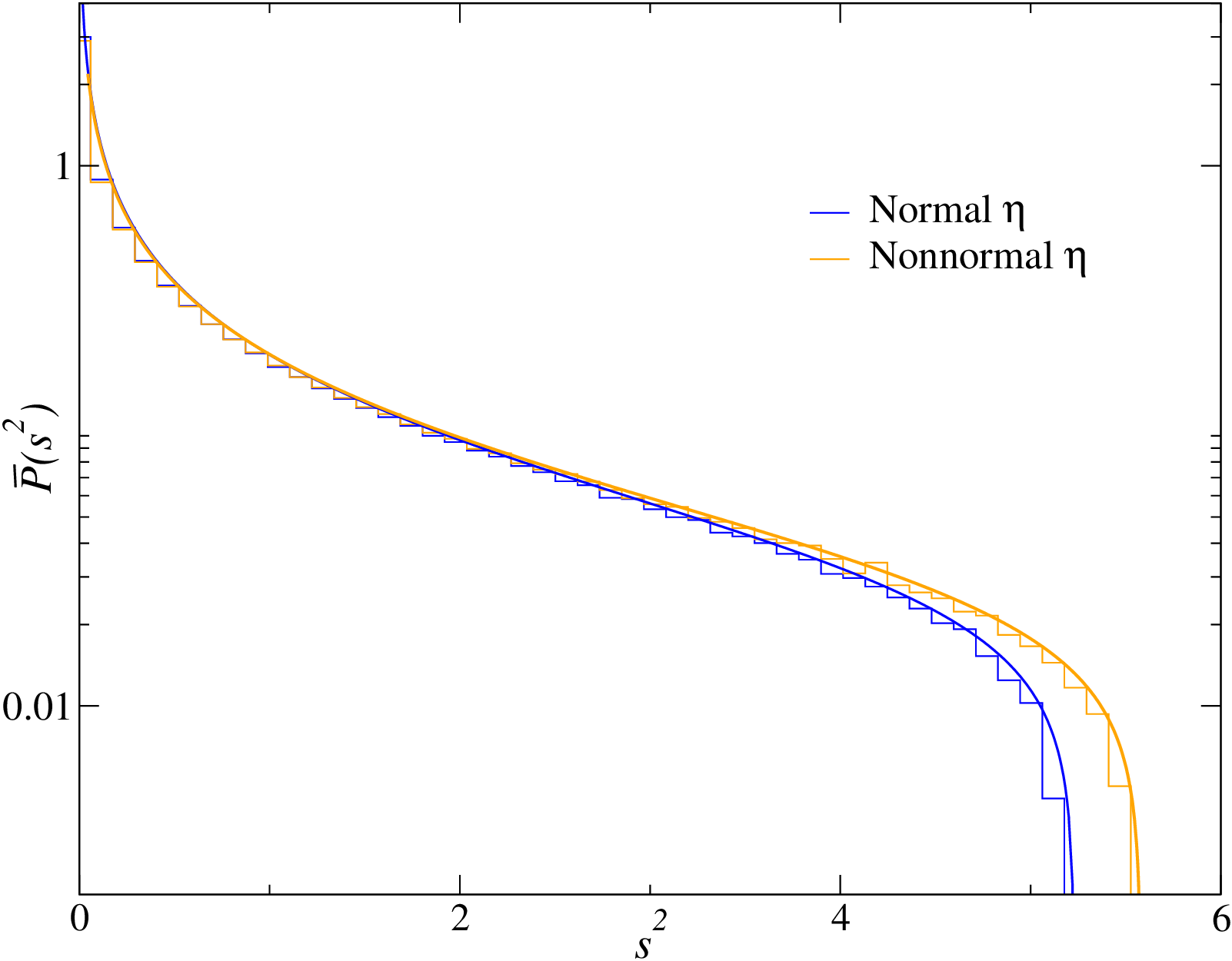}
                \caption{(Color online) Density, $\overline{P}(s^{2})$, of the square of the singular values of $\mathsf{C}$ where $s$ represent the singular values. Here the $\eta$ corresponds to the case ($v$) where the same values for the parameters $c,\,p$ and $q$ have been used as in Fig. \ref{nonsym}. Using solid lines in figure we plot the theory \cite{vin2013} which depends on the spectrum of $\eta\eta^{t}$. Histograms, shown by left-stair case, are calculated using the eigenvalues of $\mathsf{CC}^{t}$ for $100$ realizations where $N=1024$ and $T=2N$. Here the {\it orange} color represent the nonnormal $\eta$ case and the {\it blue} color represent the normal $\eta$ case. }
\label{Sval}
\end{figure}

For the second case we consider $\eta_{jk}=c\,\delta_{jk}$. Using Eqs. (\ref{a1d1}-\ref{a2d2}) and Eqs. (\ref{gLjj}-\ref{res-g22b}), we get
\begin{eqnarray}
\overline{\mathsf{a}}_{1}&=&(w-c\,\overline{g}_{22})\mathbf{1}_{N}, ~\overline{\mathsf{a}}_{2}=(w-\kappa \,c\,\overline{g}_{11})\mathbf{1}_{T},\nonumber\\
\overline{\mathsf{d}}_{1}&=&(w^{*}-c\,\overline{g}_{\overline{2}\,\overline{2}})\mathbf{1}_{N},~\overline{\mathsf{d}}_{2}=(w^{*}-\kappa \,c\,\overline{g}_{\overline{1}\,\overline{1}})\mathbf{1}_{T}.
\end{eqnarray}
Next, from Eqs. (\ref{gam11}-\ref{gam22}) we get
\begin{eqnarray}
\langle \overline{\Gamma}^{(1)}_{1} \rangle&=& \langle \overline{\Gamma}^{(1)}_{2} \rangle=\left\langle\frac{1}{\overline{\mathsf{a}}_{1}\,\overline{\mathsf{d}}_{1}-\overline{g}_{2\overline{2}}\overline{g}_{\overline{2}2}\mathbf{1}_{N}}\right\rangle \equiv\overline{\gamma}_{1},\\
\langle \overline{\Gamma}^{(2)}_{1} \rangle&=& \langle \overline{\Gamma}^{(2)}_{2} \rangle=\left\langle\dfrac{1}{\overline{\mathsf{a}}_{2}\,\overline{\mathsf{d}}_{2}-\kappa^{2}\overline{g}_{1\overline{1}}\overline{g}_{\overline{1}1}\mathbf{1}_{T}}\right\rangle\equiv\overline{\gamma}_{2}.
\end{eqnarray}
Using these in Eqs. (\ref{res1-gjj}-\ref{res-g22b}), the set of equations we obtain is 
\begin{eqnarray}\label{gdigam}
\overline{g}_{jj}=\overline{\mathsf{d}}_{j}\overline{\gamma}_{j},
~\overline{g}_{\overline{j}\,\overline{j}}=\overline{\mathsf{a}}_{j}\overline{\gamma}_{j},
\end{eqnarray}
{where $j=1$ and $2$,} and 
\begin{eqnarray}
\label{spec-Ave}
\overline{g}_{1\,\overline{1}}=\overline{g}_{2\,\overline{2}}\overline{\gamma}_{1},~
\overline{g}_{2\,\overline{2}}=\kappa\overline{g}_{1\,\overline{1}}\overline{\gamma}_{2},\nonumber\\
\overline{g}_{\overline{1}1}=\overline{g}_{\overline{2}2}\overline{\gamma}_{1},~
\overline{g}_{\overline{2}2}=\kappa\overline{g}_{\overline{1}1}\overline{\gamma}_{2}.
\end{eqnarray}
Finally, the identity (\ref{Id-gam1gam2}) gives
\begin{equation}\label{id-etad}
\overline{\gamma}_{1}\overline{\gamma}_{2}=\frac{1}{\kappa}.
\end{equation}
\begin{figure}
        \centering
               \includegraphics [width=0.45\textwidth]{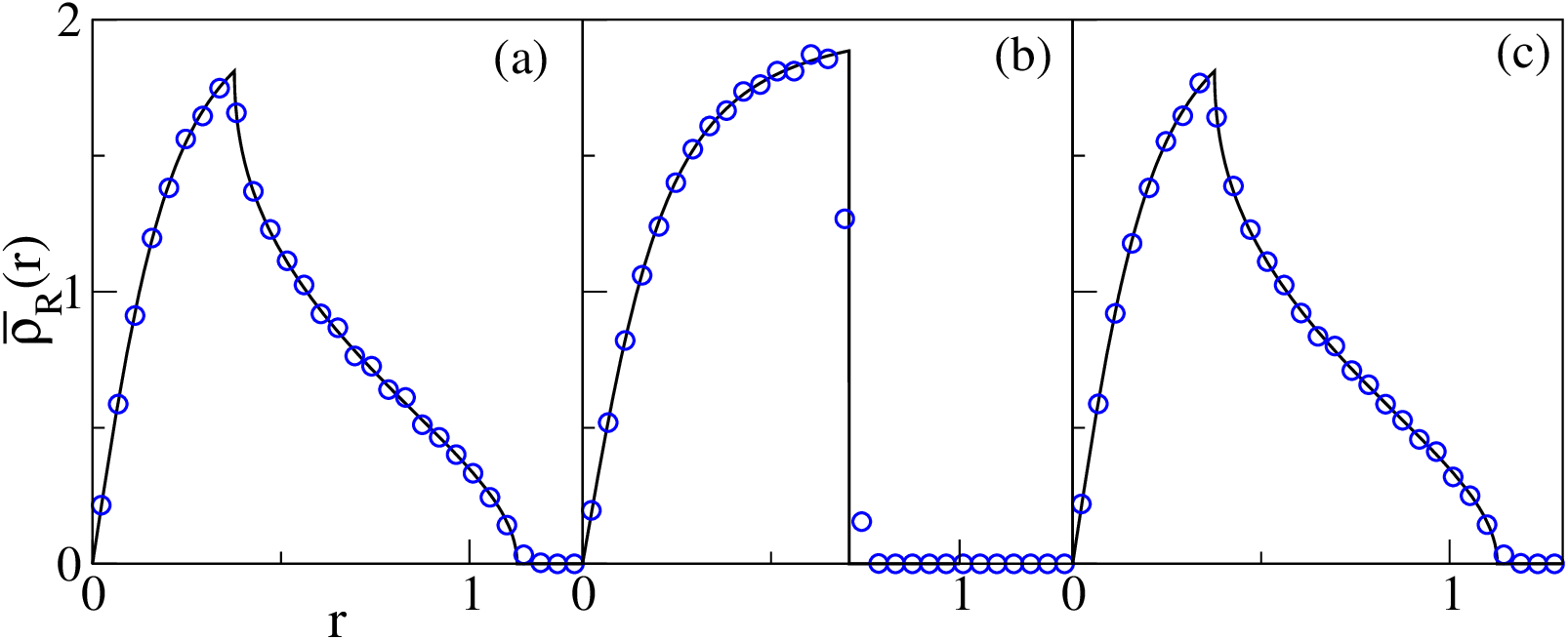}
                \caption{(Color online) Radial distribution, $\overline{\rho}_{\mathsf{R}}(r)$, of the densities of absolute values of the {eigenvalues} of $\mathsf{C}$ corresponding the $\eta_{jk}=c\delta_{jk}$ case. In this figure we show results for $c=-1/4,\,0$ and $1/4$ respectively in (a), (b) and (c). The corresponding scatter plots of these are shown respectively in Fig. \ref{dif-d}(a), (b) and (c). Theory for $\overline{\rho}_{\mathsf{R}}(r)$ is obtained 
{from Eq. (\ref{ellipse}) using} $z=r\,\exp(\ii \theta)$. In this figure the matrix dimension $N=512$ and we have used $300$ independent realizations to obtain {data shown by open circles}.}
\label{NFig7}
\end{figure}

In the holomorphic region, these equations have a simple solution, viz., $\overline{g}_{j\overline{j}}=\overline{g}_{\overline{j}j}=0$. In the nonholomorphic region, which is described by the ellipse (\ref{ellipse}), we solve this equations by observing a linear relation between the $\overline{\gamma}_{j}$'s. However, to be consistent with the notation used in Ref. \cite{Burda:2010}, we instead use $\overline{\mu}_{j}$ for $|w|^{2}\overline{\gamma}_{j}-1$, for $j=1,2$. 
{To obtain the linear relation we first write $\overline{g}_{2\overline{2}}\overline{g}_{\overline{2}2}$ in terms of $\overline{g}_{1\overline{1}}\overline{g}_{\overline{1}1}$ using Eq. (\ref{spec-Ave}) and then we use the identity (\ref{id-etad}). This method leads to} a linear relation between the $\overline{\mu}_{j}'$s:
\begin{equation}\label{mu1mu2}
\overline{\mu}_{2}=-c^{2}(1-\kappa)+\kappa\overline{\mu_{1}}.
\end{equation}

Note that together with the identity (\ref{id-etad}), the above relation simplifies the problem to a quadratic equation since the Green's function, we wish to calculate inside the ellipse (\ref{ellipse}), is given by
\begin{equation}
\overline{\mathbf{g}}_{\mathsf{C}}(z,z^{*})=\frac{\kappa\overline{\mu}_{1}+\overline{\mu}_{2}+(\kappa+1)\,c^{2}-2cz+2(1-c^{2})\kappa}{2(1-c^{2})\kappa z}.
\end{equation}
The above relation is obtained using the Eq. (\ref{MC-MP}) and the solutions of the resulting self-consistent equations for the $\overline{g}_{jj}$'s; see (\ref{res1-gjj}-\ref{res-g22b}). There we find
\begin{equation}
\overline{g}_{11}=\frac{w^{*}\overline{\gamma}_{1}-c\,w/\kappa}{1-c^{2}},~\overline{g}_{22}=\frac{w^{*}\overline{\gamma}_{2}-c\,w}{1-c^{2}}.
\end{equation}

\begin{figure}
        \centering
               \includegraphics [width=0.45\textwidth]{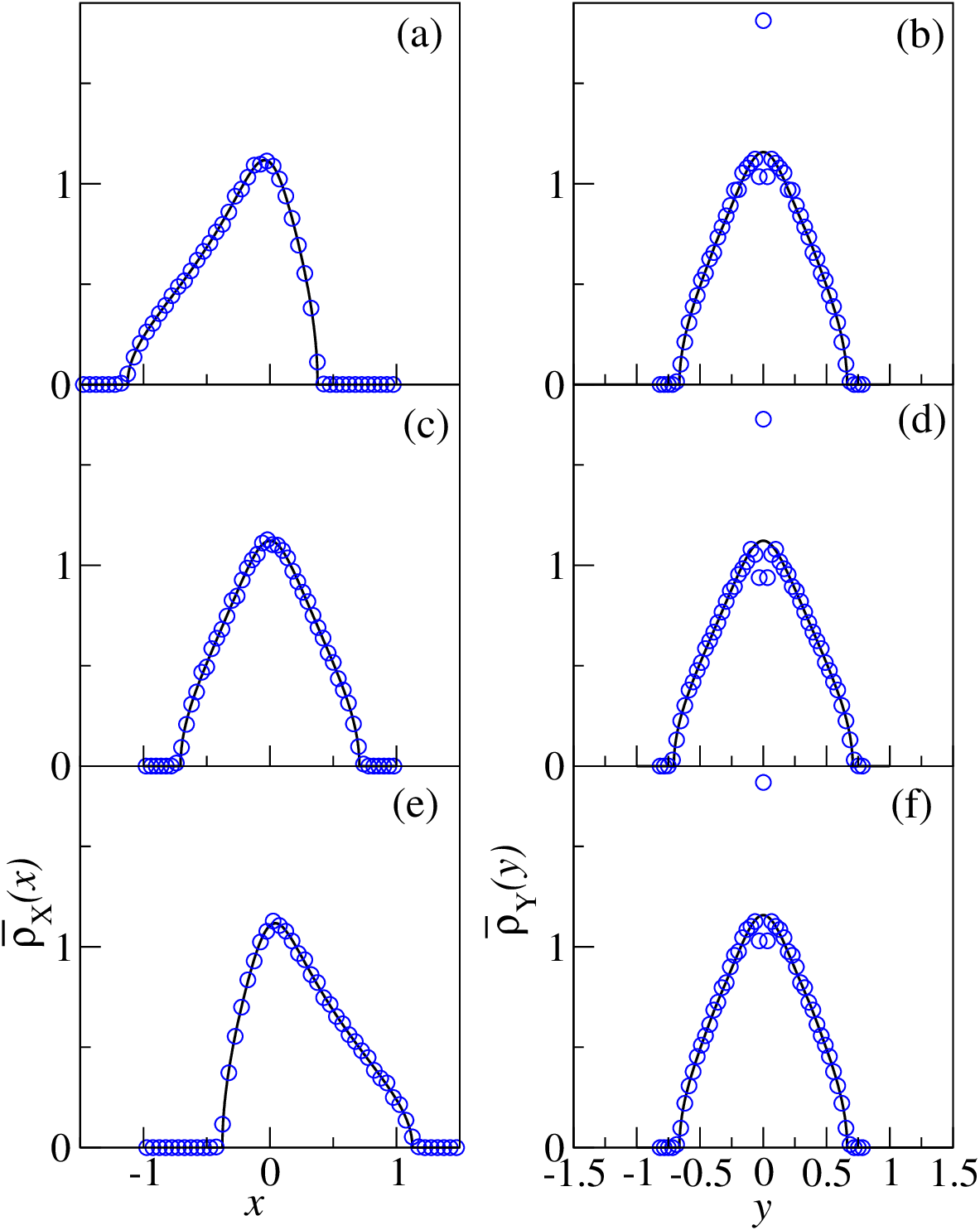}
                \caption{(Color online) Marginal distributions $\overline{\rho}_{X}(x)$, shown in the left panel, and $\overline{\rho}_{Y}(y)$, {shown in the right panel}, describing respectively the densities of real and imaginary part of the eigenvalues {of $\mathsf{C}$ where $\eta_{jk}=c\delta_{jk}$}. In this figure we {show results for} $c=-1/4,\,0$ and $1/4$ from top to bottom. The corresponding scatter plots of these are shown respectively in Fig. \ref{dif-d}(a), (b) and (c). {Theory for $\overline{\rho}_{X}(x)$ and $\overline{\rho}_{Y}(y)$ is obtained from Eq. (\ref{d-dens}) using $z=x+\ii y$}. In this figure the matrix dimension $N=512$ where we have used $300$ independent realizations to obtain {data shown by open circles.}}
\label{d-den}
\end{figure}
{We re-write Eq. (\ref{id-etad}) for the $\overline{\mu}_{j}$'s and solve it using Eq. (\ref{mu1mu2}). We obtain}
\begin{eqnarray}\label{gamma1n2}
\overline{\mu}_{1}+1
&=&
\frac{-\alpha(1-\kappa)+\sqrt{\alpha^{2}(1-\kappa)^{2}+4|w|^{4}}}{2\kappa},\nonumber\\
\overline{\mu}_{2}+1
&=&
\frac{\alpha(1-\kappa)+\sqrt{\alpha^{2}(1-\kappa)^{2}+4|w|^{4}}}{2},
\end{eqnarray}
where $\alpha=(1-c^{2})$. Consequently we derive the Green's function:
\begin{equation}
\overline{\mathbf{g}}_{\mathsf{C}}(z,z^{*})=
\frac{\alpha(\kappa-1)+\sqrt{\alpha^{2}(1-\kappa)^{2}+4|z|^{2}}}{2\alpha\kappa z}-\frac{c}{\kappa\alpha}.
\end{equation}
Finally, using the Gauss law (\ref{Gauss}), for $\kappa\le1$, we find 
\begin{equation}\label{d-dens}
\overline{\rho}_{\mathsf{C}}(z,z^{*})=
\frac{1}{\pi \kappa\alpha\sqrt{\alpha^{2}(1-\kappa)^{2}+4|z|^{2}}}̣̣̣̣.
\end{equation}
On the other hand, the density is zero outside the ellipse (\ref{ellipse}). Taking account the $1/z$ terms for $z\rightarrow0$, and the identity $\partial/\partial z^{*} (1/z)=\delta(z)\delta(z^{*})$, we get an additive term $(1-\kappa^{-1})\delta(z)\delta(z^{*})$ for $\kappa>1$. This term is consistent with the note on the rank of $\mathsf{C}$; see Sec. \ref{General}. For $c=0$, this result (\ref{d-dens}) yields the spectral density for the uncorrelated case which matches with the result obtained in Ref. \cite{Burda:2010} if we replace $z$ by $\sqrt{\kappa}\,z$.

\begin{figure}
        \centering
               \includegraphics [width=0.45\textwidth]{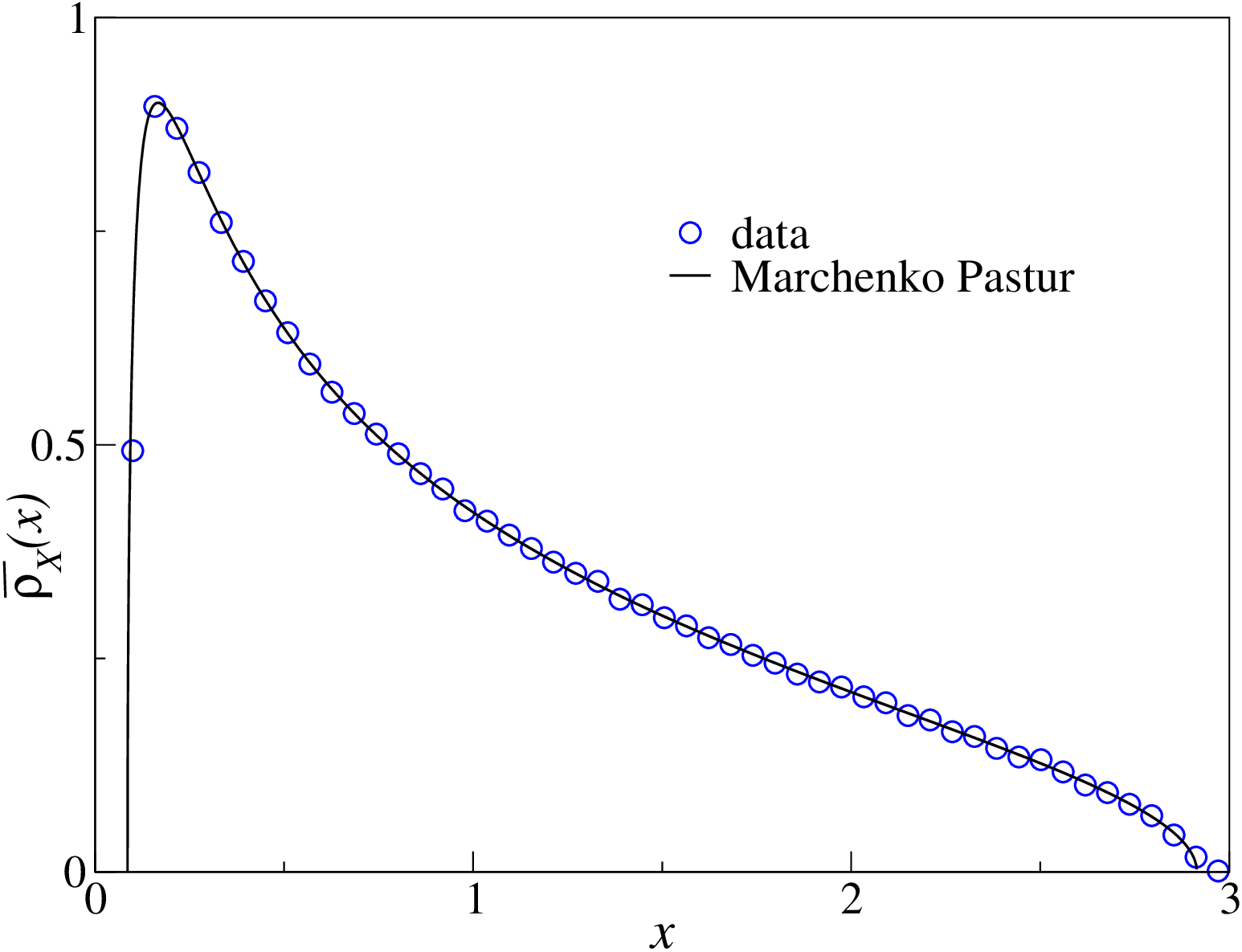}
                \caption{(Color online) Marginal density $\overline{\rho}_{X}(x)$ of {the real part of the eigenvalues of $\mathsf{C}$ for} $\eta_{jk}=c\delta_{jk}$ where $c=0.9999$. The histogram, shown by open circles, is calculated from $225$ realizations of $512$-dimensional $\mathsf{C}$. The Mar\v cenko Pastur density is shown with the solid line. As in all previous figures, here as well, we have considered $T=2N$.}
\label{xden}
\end{figure}

In Fig. \ref{NFig7}, we compare the radial distribution, $\overline{\rho}_{\mathsf{R}}(r)$, obtained {from} our theory (\ref{d-dens}) {using} $z=r\,\exp(\ii \theta)$. We consider three different $c$ values as chosen in Fig. \ref{dif-d}. For instance in Fig. \ref{NFig7}(a) we compare the radial density with the theory (\ref{d-dens}) for $c=-1/4$. Since the radial density does not depend on the sign of $c$, in figure (c) we get the same result for $c=1/4$. The finite-$N$ effects, as can be seen for $c=0$ in figure (b) \cite{Burda:2010, Kanzieper-Singh}, are in the peaks and in the tails of the densities. Similarly in Fig. \ref{d-den}, we compare {the} marginal density distribution of the real and the imaginary parts of the eigenvalues which is obtained by integrating over the variable other than the variable which distribution is sought for using $z=x+\ii y$. For example, the density for the real part, $\overline{\rho}_{X}(x)$, is obtained by integrating over $y$ from $\overline{\rho}_{X,Y}(x,y)\equiv\overline{\rho}(z,z^{*})$. For $c=0$, both densities are symmetric along both axes. However, for $c\ne0$, the symmetry along the $x$-axis is broken. As shown in the figure, our theory closely describes these densities except the peaks in $\overline{\rho}_{Y}(y)$ in between the dips near $y=0$. These in fact are finite-$N$ effects. These peaks are well studied for the Ginibre orthogonal ensemble \cite{Ginibre} and for uncorrelated matrices where analytical results are known in much detail \cite{Kanziper, Adelman, Forrester}. 

An interesting consequence for  $\overline{\rho}_{X}(x)$ is observed from a analytic simple calculation using the result (\ref{d-dens}). It can be shown that as $c\to1$, $\overline{\rho}_{X}(x)$ converges to the Mar\v cenko Pastur distribution \cite{marchenko} implying thereby that on average the lowest of the real components of the spectra is non-negative. To demonstrate this theory we consider $c=0.9999$ and compare $\overline{\rho}_{X}(x)$ with the Mar\v cenko Pastur result in Fig. \ref{xden}. For this value of $c$, we get $\sim 2-3\%$ nonzero $y$-components which are of $O(x_{+})$ where $x_{+}$ is the upper edge of $\overline{\rho}_{X}(x)$. However, as {it} can be seen in this figure, the Mar\v cenko Pastur result reasonably describes $\overline{\rho}_{X}(x)$.

\section{Summary and Conclusion}\label{Conc}
To summarize the work, we begin with a few important remarks. In physics, nonsymmetric random matrices have been of importance from different perspectives \cite{Brody81, GuhrGW98, BenRMP97, NHVerbaarschot}. For the correlation matrices, however, these are important in order to understand time evolution of a multivariate complex system or to understand mutual behaviour of two different statistical systems. Usually, Wishart's approach is used in dressing the noise from an empirical correlation matrix where the spectral analysis {plays a crucial role. Theory incorporating linear} correlations, however, provides a better way of understanding effects of the actual correlations in the spectral statistics. The importance is evident from Refs. \cite{Finance1, Finance2, Finance3, Finance4, vrt2013, guhr:2013}. From the theoretical viewpoint, there remain many problems like obtaining the finite-$N$ results where some methods developed for the correlated Wishart ensembles \cite{guhr1, MousSimon} seem potentially useful. 

In this paper we have focused mainly on the domain of the bulk of complex eigenvalues under the influence of actual correlations. We have derived analytic result for the contour enclosing {the} bulk of the eigenvalues for a general $\eta$ for which the full correlation matrix $\xi$ is positive definite. One important finding {here} is the ellipse which explains {the} boundary of eigenvalues for a system dominated by autocorrelations of equal strength. Beside that we have shown that the contour depends on the non-normality of the correlation matrix. In this paper, the results are illustrated using tridiagonal $\eta$ matrices, which {are special but allows us to obtain analytical results, which may well be used in applications}. Also we have remarked that effects of the non-normality can also be observed in the singular values of $\eta$, mostly for small or large eigenvalues. 

As far as applications are concerned, perhaps the closest example is in the quantitative finance where study of the correlations between stock prices with their volume data \cite{Poland} could be good starting point. However, the main emphasis has so far been the bi-variate time series \cite{Boris}. We believe that our results can be used for the multi-stock data analysis which may shed more light on the non-trivial behavior of the financial market. Finally, we believe that our results may not be confined to the multivariate analysis as can be deduced from the historical developments and applications of RMT \cite{GuhrGW98,SFV}. 

\section{Acknowledgments}
The authors are grateful to Thomas H. Seligman and F. Leyvraz for useful discussions and consistent encouragements. Financial support from CONACyT through Project No. 154586 and No. PAPIIT UNAM RR 114014 is acknowledged. Vinayak was supported by DGAPA/UNAM as a postdoctoral fellow. 

\appendix

\section{Inverse {of the matrix $\tilde{\mathsf{W}}-\Sigma$}}\label{apInv}
The inverse of matrix we want to calculate may be viewed as
\begin{equation}
\left(\tilde{\mathsf{W}}-\Sigma
\right)\equiv \mathsf{M}=
\left(
\begin{matrix}
\mathbf{a} & \mathbf{b}\\
\mathbf{c} & \mathbf{d}
\end{matrix}
\right),
\end{equation}
where all the matrices are of dimension $(N+T)\times (N+T)$. Matrices $\mathbf{b}$ and $\mathbf{c}$ are diagonal:
\begin{eqnarray}
\mathbf{b}&=&
-\left(
\begin{matrix}
\overline{g}_{2\,\overline{2}}\mathbf{1}_{N}& \mathbf{0}\\
\mathbf{0}& \kappa\overline{g}_{1\,\overline{1}}\mathbf{1}_{T}
\end{matrix}
\right),~
\mathbf{c}=
-\left(
\begin{matrix}
\overline{g}_{\overline{2}\,2}\mathbf{1}_{N}& \mathbf{0}\\
\mathbf{0}& \kappa\overline{g}_{\overline{1}\,1}\mathbf{1}_{T}
\end{matrix}
\right),\nonumber\\
\end{eqnarray}
and for the diagonal blocks we have
\begin{eqnarray}
\mathbf{a}=\left(
\begin{matrix}
\overline{\mathsf{a}}_{1}& \mathbf{0}\\
\mathbf{0}& \overline{\mathsf{a}}_{2}
\end{matrix}
\right),~
\mathbf{d}=\left(
\begin{matrix}
\overline{\mathsf{d}}_{1}& \mathbf{0}\\
\mathbf{0}& \overline{\mathsf{d}}_{2}
\end{matrix}
\right).
\end{eqnarray}
We use Schur components of $\mathbf{a}$ and of $\mathbf{d}$, viz., $\mathbf{d}-\mathbf{c}\mathbf{a}^{-1}\mathbf{b}$ and $\mathbf{a}-\mathbf{b}\mathbf{d}^{-1}\mathbf{c}$, respectively, assuming that none of them are singular. Then the inverse of the matrix \cite{Bernstien} we get is
\begin{equation}
 \mathsf{M}^{-1}
=\left(
\begin{matrix}
\mathsf{S}^{-1}& -\mathsf{S}^{-1}\mathbf{b}\mathbf{d}^{-1}\\
-\mathbf{d}^{-1}\mathbf{c}\mathsf{S}^{-1}& (\mathbf{d}-\mathbf{c}\mathbf{a}^{-1}\mathbf{b})^{-1}
\end{matrix}
\right),
\end{equation}
where $\mathsf{S}=\mathbf{a}-\mathbf{b}\mathbf{d}^{-1}\mathbf{c}$. Thus we get (\ref{DSF})


\begin{thebibliography}{99}

\bibitem{bachelier} L. Bachelier, Theorie de la Speculation, {\it The Random Character of Stock Market Prices},  Gauthier-Villars, Paris 1964, Pages: 17-78.

\bibitem{Wilks}   S. S. Wilks, {\it Mathematical Statistics} (Wiley, New York, 1962).

\bibitem{Muirhead}
R.~J. Muirhead, \emph{Aspects of Multivariate Statistical Theory}, Wiley
  Series in Probability and Statistics (Wiley, 2009).

\bibitem{Wishart}
J.~Wishart, Biometrika {\bf 20A}, 32 (1928).


\bibitem{Finance1}
L.~Laloux, P.~Cizeau, J.~P.~Bouchaud, and M.~Potters, Phys. Rev. Lett. {\bf 83}, 1467 (1999).

\bibitem{Finance2}
V.~Plerou, P.~Gopikrishnan, B.~Rosenow, L.~A.~N.~Amaral, and H.~E.~Stanley, Phys. Rev. Lett. {\bf 83}, 1471 (1999).

\bibitem{Finance3}
V.~Plerou, P.~Gopikrishnan, B.~Rosenow,  L.~A.~Nunes~Amaral, T.~Guhr, and H.~E.~Stanley, Phys. Rev. E {\bf65} 066126 (2002).

\bibitem{Finance4}
F.~Lillo and R.~N.~Mantegna, Phys. Rev. E {\bf 72}, 016219 (2005).

\bibitem{Thomas2012}
M.~C.~M\"{u}nnix, T.~Shimada, R.~Sch\"{a}fer, F.~Leyvraz, T.~H.~Seligman, T.~Guhr, and H.~E.~Stanley,  Sci. Rep. {\bf 2}, 644 (2012).


\bibitem{gene}
F.~Luo, J.~Zhong, Y.~Yang, and J.~Zhou, Phys. Rev. E {\bf 73} 031924 (2006).

\bibitem{Seba:03}
P.~\v Seba, Phys. Rev. Lett. {\bf 91}, 198104 (2003).

\bibitem{diverseAT}
M.~S.~Santhanam and P.~K.~Patra, Phys. Rev. E {\bf 64}, 016102 (2001).

\bibitem{Mehta}   M. L. Mehta, {\it Random Matrices} (Academic Press, New York, 2004).

\bibitem{Brody81}
 T.~A.~Brody, J.~Flores, J.~B.~French, P.~A.~Mello, A.~Pandey, and S.~S.~M.~Wong, Rev. Mod. Phys. {\bf 53}, 385 (1981).

\bibitem{GuhrGW98}
T.~Guhr, A.~M.~Groeling and H.~A.~Weidenm\"{u}ller, Phys. Rep. {\bf299}, {189} (1998).

\bibitem{BenRMP97}
C.~W.~J.~Beenakker, Rev. Mod. Phys. {\bf 69}, 731 (1997).

\bibitem{Verbaarschot}  T.~Nagao and M.~Wadati, J. Phys. Soc. Jpn {\bf 60}, 3298 (1991); J.~J.~M.~Verbaarschot and I.~Zahed, Phys. Rev. Lett. {\bf 70}, 3852 (1993); J.~J.~M.~Verbaarschot, Nucl. Phys. B {\bf 426} 559 (1994).

\bibitem{Muller:Review}
R.~R.~M\"{u}ller, IEEE Trans. Inf. Theory, {\bf 48}, 2495 (2002);

\bibitem{marchenko}
V.~A.~Mar\v cenko and L.~A.~Pastur, Math.~USSR~Sb. {\bf 1}, 457 (1967).


\bibitem{vrt2013} 
Vinayak, R.~Sch\"{a}fer, and T.~H.~Seligman, Phys. Rev. E {\bf 88}, 032115 (2013).


\bibitem{Silverstien}  J. W. Silverstein, J. Multivariate Anal. {\bf 55}, 331 (1995).

\bibitem{SenM}
A.~ M.~Sengupta and P.~P.~Mitra, Phys. Rev. E {\bf 60}, 3389 (1999).


\bibitem{Burda:2005}
Z.~Burda, J.~Jurkiewicz, and B.~Waclaw, Phys. Rev. E {\bf 71}, 026111 (2005).


\bibitem{MousSimon}
S.~H.~Simon and A.~L.~Moustakas, Phys. Rev. E {\bf 69}, 065101(R) (2004).
\bibitem{Baik:2005} J.~Baik\etal{, G.~Ben Arous, and S.~P\' ech\' e} {\it Ann.~Probab.} {\bf 33} 1643 (2005)
; S.~P\'{e}ch\'{e} 
{\it J.~Multivariate ~Anal.} {\bf 97}, 874 (2006).

\bibitem{vp2010}  
Vinayak and A.~Pandey, Phys. Rev. E {\bf 81}, 036202 (2010).


\bibitem{guhr1}
C.~Recher, M.~Kieburg, and T.~Guhr, Phys. Rev. Lett. {\bf 105}, 244101 (2010);
C.~Recher, M.~Kieburg, T.~Guhr, and M.~R.~Zirnbauer, J. Stat. Phys. {\bf 148}, 981 (2012).

\bibitem{John} I.~M.~Johnstone, Ann. Stat. {\bf 29}, 295 (2001).


\bibitem{Biely:08} 
C.~Biely and S.~Thurner, Quant. Financ. {\bf 8}, 705 (2008).

\bibitem{Kwapien:2006} 
{\bf J.~Kwapie\'n, S.~Dro\.zd\.z , A.~Z.~G\'orski, and P.~O\'swiecimka, Acta. Phys. Pol. B {\bf 37}, 3039 (2006).} 


\bibitem{Bouchaud:2007} 
J.-P.~Bouchaud, L.~Laloux, M.~A.~Miceli and M.~Potters, Eur. Phys. J. {\bf 55}, 201 (2007).

\bibitem{Bouchaud:2009}
J.~P.~Bouchaud and M.~Potters, arXiv:0910.1205.

\bibitem{Stanley:2011} 
D.~Wang, B.~Podobnik, D.~Horvati\'{c}, and H. E. Stanley, Phys. Rev. E {\bf 83}, 046121 (2011).

\bibitem{Livan:2012} 
G.~Livan and L.~Rebecchi, Eur. Phys. J. B {\bf 85}, 213 (2012).

\bibitem{Dorzdz} 
S.~Dro\.zd\.z , J.~Kwapie\'n , A.~A.~Ioannides, Acta. Phys. Pol. B {\bf 42}, 987 (2011).

\bibitem{PRDorzdz}
J.~Kwapie\'n, S.~Dro\.zd\.z, Phys. Rep. {\bf 515}, 115 (2012).

\bibitem{Kwapien:2000}
{\bf J.~Kwapie\'n, S.~Dro\.zd\.z, and A.~A.~Ioannides, Phys. Rev. E {\bf 62}, 5557 (2000).}



\bibitem{Burda:2010}
Z.~Burda, A.~Jarosz, G.~Livan, M.~A.~Nowak, and A.~Swiech, Phys. Rev. E {\bf 82}, 061114 (2010).


\bibitem{Kanzieper-Singh} 
E.~Kanzieper and N.~Singh, J. Math. Phys. {\bf 51}, 103510 (2010).


\bibitem{Proof1}
S.~O'rourke and A.~Soshnikov, arXiv:1012.4497.

\bibitem{Proof2} 
F. G\"otze, A. Tikhomirov, Electron. J. Probab. {\bf 16}, 2219 (2011).

\bibitem{Mario:2014} J.~R.~Ipsen and M.~Kieburg, Phys. Rev. E {\bf 89}, 032106 (2014).

\bibitem{vin2013} 
Vinayak, Phys. Rev. E {\bf 88}, 042130 (2013).

\bibitem{Muller}
R.~R.~M\"{u}ller, Acta Phys. Pol. B {\bf 36}, 1001 (2005).

\bibitem{Akeman:2013} 
G.~Akemann, J.~R.~Ipsen, and M.~Kieburg, Phys. Rev. E {\bf 88}, 052118 (2013).

\bibitem{Pastur}  L.~A.~Pastur, Theoret.~and~Math.~Phys.II {\bf 10}, 67 (1972).


\bibitem{Sommers}  
H.-J. Sommers, A. Crisanti, H. Sompolinsky, and Y. Stein, Phys. Rev. Lett. {\bf 60}, 1895 (1988).

\bibitem{Nowak-Nowak} 
E.~Gudowska-Nowak, R.~A.~Janik, J.~Jurkiewicz, and M.~A.~Nowak,  Nucl. Phys. B {\bf 670}, 479 (2003͒). 
 
\bibitem{Joshua-Zee} 
J.~Feinberg and A.~Zee, Nuclear Physics B {\bf 504}, 579 (1997).

\bibitem{Janik-Nowak} 
R.~A.~Janik, M.~A.~Nowak, G.~Papp, J.~Wambach, and I.~Zahed, Phys. Rev. E {\bf 55}, 4100 (1997);
R.~A.~Janik, M.~A.~Nowak, G.~Papp, and I.~Zahed, Nucl. Phys. B {\bf 501}, 603 (1997).

\bibitem{Burda-Janik:2010} 
Z.~Burda, R.~A.~Janik, and B.~Waclaw, Phys. Rev. E {\bf 81}, 041132 (2010).

\bibitem{Weyl} 
H.~Weyl, Proc. Nat. Acad. Sci. U. S. A. {\bf 35}, 408 (1949); 
A.~Horn, Proc. Amer. Math. Soc. {\bf 5}, 4 (1954).

\bibitem{Brezin} E.~Brezin, C.~Itzykson, G.~Parisi, and J.-B.~Zuber, Commun.~Math.~Phys. {\bf 59}, 35 (1978).

\bibitem{ap81}  A.~Pandey, Ann. Phys. (N.Y.) {\bf 134}, 110 (1981).

\bibitem{Wilkinson} J.~H.~Wilkinson {\it The perfidious polynomial}, Studies in Numerical Analysis, ed. by G. H. Golub, pp. 1–28. (Studies in Mathematics, {\bf 24}, 1984, Washington, D.C.: Mathematical Association of America); L.~N.~Trefethen and D.~Bau (1997), Numerical Linear Algebra SIAM (1997).

\bibitem{Kulk} 
D.~Kulkarni, D.~Schmidt and Sze-Kai~Tsui, Linear Algebra and its Applications {\bf 297}, 63 (1999).


\bibitem{bulk} It is understood that in both cases the bulk density is normalized to $1-N^{-1}$. 

\bibitem{vmarko}
Vinayak and M.~\v Znidari\v c, J. Phys. A: Math. Theor. {\bf 45},  125204 (2012).

\bibitem{Ginibre} 
J.~Ginibre, J. Math. Phys. (N.Y.) {\bf 6}, 440 (1965).


\bibitem{Kanziper}
E.~Kanzieper and G.~Akemann, Phys. Rev. Lett. {\bf 95}, 230201 (2005); E.~Kanzieper and G.~Akemann, J. Stat. Phys. {\bf 129}, 1159 (2007).
\bibitem{Forrester} 
P.~J.~Forrester and T.~Nagao, Phys. Rev. Lett. {\bf 99}, 050603 (2007); P.~J.~Forrester, arXiv:1309.7736.
\bibitem{Adelman} 
A.~Edelman and N.~R.~Rao, Acta Numerica {\bf 14}, 233 (2005).


\bibitem{NHVerbaarschot} 
M.~A.~Halasz, J.~C.~Osborn, and J.~J.~M.~Verbaarschot, Phys. Rev. D {\bf 56}, 7059 (1997).


\bibitem{guhr:2013} 
T.~Wirtz and T.~Guhr, Phys. Rev. Lett. {\bf 111}, 094101 (2013).

\bibitem{Poland}  
K.~Karpio , P.~$\L$ukasiewicz, and A.~Orlowski, Acta. Phys. Pol. A {\bf 121}, b61 (2010).


\bibitem{Boris} 
B.~Podobnik and H.~E.~Stanley, Phys. Rev. Lett. {\bf 100}, 084102 (2008).

\bibitem{SFV}
N.~C.~Snaith, P.~J.~Forrester, and J.~J.~M.~Verbaarschot, J.~Phys.~A {\bf 36}, R1-R10, 2859 (2004).

\bibitem{Bernstien} D.~S.~Bernstein {\it Matrix Mathematics: Theory, Facts and Formulas}, (Princeton University Press, 2005).

\end{thebibliography}
\end{document}